\documentclass[11pt]{article}
\usepackage{amsfonts}
\usepackage{mathrsfs}
\usepackage{hyperref}
\usepackage{subfigure}
\usepackage{amsmath}
\usepackage{amssymb}
\usepackage{bbold}
\usepackage{geometry}
\usepackage{longtable}
\usepackage{cite}
\usepackage{lscape,graphicx} 
\usepackage{graphics}
\usepackage{color}
\usepackage[numbers,sort&compress]{natbib}

\newcommand{\trace}[1]{\ensuremath{\mathrm{Tr}#1}}

\begin{document}

{\small
\begin{flushright}
CP3-Origins-2018-47 DNRF90 \\[30pt]
\end{flushright} }

\begin{center}
{\Large \bf  Safe Trinification }\\[20pt]
Zhi-Wei Wang$^{1,2}$~Abdulrahim Al Balushi$^{1}$~Robert Mann$^{1}$~Hao-Miao Jiang$^{1}$\\[10pt]
{\it \small
$^{1}$Department of Physics, University of Waterloo, Waterloo, On N2L 3G1, Canada\\
 $^{2}$CP$^3$-Origins, University of Southern Denmark,\\
Campusvej 55, DK-5230 Odense M, Denmark\\[4pt]
}

\begin{abstract}
In this work, we provide a UV safe Trinification theory in which the Standard Model is embedded. Using   recently developed large number-of-flavor techniques, safety is achieved by adding to the theory gauged vector-like fermions.  We find that all gauge, scalar quartic, and Yukawa couplings achieve an interacting ultraviolet fixed point below the Planck scale.  We find renormalization group flow solutions matching the Standard Model in the IR,  indicating a truly UV completion of the Standard Model.  Imposing constraints that realistic Higgs, top, bottom, tau and reasonable neutrino masses are recovered, we find the set of allowed solutions to be quite restrictive. 
Furthermore, the symmetry breaking scale is predicted to
be around 10 TeV, making this model vulnerable to experiment.   
\end{abstract}

\end{center}

\section{Introduction}
According to Wilson \cite{Wilson:1971bg,Wilson:1971dh}, a theory is fundamental if it features an ultraviolet (UV) fixed point which is either non-interacting (asymptotically free) \cite{Gross:1973ju,Cheng:1973nv,Callaway:1988ya,Giudice:2014tma,Holdom:2014hla,Pica:2016krb,Molgaard:2016bqf,Gies:2016kkk,Einhorn:2017jbs,Hansen:2017pwe} or interacting (asymptotically safe) \cite{Litim:2014uca,Litim:2015iea,Esbensen:2015cjw}.
The first indisputable and precisely calculable example of a four dimensional, non supersymmetric, complete asymptotically safe quantum field theory without gravity was discovered in \cite{Litim:2014uca}. Recently, starting with the conjecture of a {\it safe} rather than {\it free}  QCD \cite{Sannino:2015sel}, the first implementation of large $N_F$ (number of flavours) technique \cite{PalanquesMestre:1983zy,Gracey:1996he,Holdom:2010qs,Pica:2010xq,Shrock:2013cca} to the whole Standard Model (with summation only in the gauge couplings) is studied in \cite{Mann:2017wzh} opening the way to the various safe extensions of the Standard Model \cite{Mann:2017wzh,Abel:2017rwl,Pelaggi:2017abg,McDowall:2018tdg,Ipek:2018sai,Molinaro:2018kjz}. Later on, extension of the large $N_F$ summation to both the gauge and Yukawa couplings is studied in \cite{Kowalska:2017pkt}, a gauge-less study with only Yukawa summation appeared in \cite{Alanne:2018ene} and for the first time to all the couplings in \cite{Pelaggi:2017abg,Antipin:2018zdg} including the semi-simple gauge group in \cite{Antipin:2018zdg}. These studies have led, for example, to enrich the original conformal window \cite{Sannino:2004qp,Dietrich:2006cm}, reviewed in \cite{Sannino:2009za,Pica:2017gcb}, with a novel asymptotically safe region ({\it conformal window 2.0}) in \cite{Antipin:2017ebo}.

At the Grand Unified Theory (GUT) frontier, the first asymptotically safe Pati-Salam model by using the large number of flavour dynamics is studied in \cite{Molinaro:2018kjz} providing a realistic possibility that the Standard Model can be UV complete. The gauge, the Yukawa and scalar quartic couplings are unified by a dynamical rather than a symmetry principle.
In this work, by using the acquired knowledge of the large $N_F$ technique, we construct a novel {\it safe Trinification} extension by adding vector-like fermions and showing that all couplings acquire a UV fixed point at energies that are far below the Planck scale. The Trinification model was first proposed in \cite{Y. Achiman,Glashow,Babu:1985gi}, and for more recent studies see e.g.~\cite{Giudice:2014tma,Pelaggi:2015kna,Sayre:2006ma,Cauet:2010ng,Hetzel:2015bla}.
The separation of scales between the UV fixed point and the Planck scale allow us to study the physics around UV fixed point while ignoring the gravitational corrections.  
The interplay with gravity has been investigated in several recent works ~\cite{Eichhorn:2018vah,Eichhorn:2017muy,Reichert:2017puo,Eichhorn:2017sok,Eichhorn:2017lry} and it will not be considered here. Differently from the usual Grand Unified scenarios \cite{Georgi:1974sy} in which only the gauge couplings unify because of their embedding into a larger group structure and then they eventually become free, in the present scenario  we have that Yukawa and scalar self couplings are intimately linked because of the safe dynamics with their high energy behavior tamed by the presence of an interacting fixed point.

The paper is organized as follows: In section \ref{Sec2} we review and introduce the Trinification \cite{Pati:1974yy} extension of the SM. In section \ref{Sec3}, we construct the minimal vector-like extension able to support a safe scenario. We analyze and classify the UV fixed point structure of the model and have also developed the large $N_F$ improved renormalization group (RG) equations and determined the couplings' evolution. In section \ref{Sec4}, we provide a detailed analysis to show the possiblity of the {\it safe Trinification} model to match the SM at IR. We offer our conclusions in section \ref{Sec5}. In appendix~\ref{App1} we summarize the one-loop RG equations for the Trinification model investigated here.

\section{Trinification extension of the Standard Model}\label{Sec2}
Consider the time-honored Trinification gauge symmetry group $G_\text{TR}$ \cite{Y. Achiman,Glashow,Babu:1985gi}.
\begin{equation}
	G_\text{TR} = SU(3)_C\otimes SU(3)_{L} \otimes SU(3)_{R}\,,
\end{equation}
with gauge couplings $g_3$, $g_L$ and $g_R$, respectively. Here $SU(3)_C$ denotes the SM color gauge group. The gauge couplings in Trinification model $\left(g_c,\,g_L,\,g_R\right)$ and the ones in the Standard Model $\left(g_3,\,g_2,\,g_Y\right)$ are related by: 
\begin{eqnarray}
    &&	g_c =g_3\,,	\quad\quad   g_{L} =g_2\,,\quad\quad g_R=\frac{2g_2g_Y}{\sqrt{3g_2^2-g_Y^2}}.\label{matching_1}
\end{eqnarray}

Compared with Pati-Salam theory, the SM quark and lepton fields are ``not'' unified into the $G_{\rm TR}$ irreducible representations. However, this disadvantage becomes an advantage in the sense that it is easier to realize the quark/lepton mass splitting. The coloured matter content of the minimal Trinification model is given by:
\begin{eqnarray}
\begin{split}\label{fermionsLR}
	\psi_{Q_{L}} &= \left(\begin{array}{ccc} u_L^1  & \mathscr{D}_L^1 & \mathscr{D}_L'^1 \\ u_L^2  & \mathscr{D}_L^2 & \mathscr{D}_L'^2\\u_L^3  & \mathscr{D}_L^3 & \mathscr{D}_L'^3\end{array}\right) \sim  (3,3,1) \,, \\ 
	\psi_{Q_{R}} &= \left(\begin{array}{ccc} u_R^1  & u_R^2 & u_R^3\\ \mathscr{D}_R^1  & \mathscr{D}_R^2 & \mathscr{D}_R^3\\\mathscr{D}_R'^1  & \mathscr{D}_R'^2 & \mathscr{D}_R'^3\end{array}\right) \sim (3,1,3) \,,  
\end{split}
\end{eqnarray}
where $i=1,2,3$ is a colour index and $\mathscr{D'}$ denotes as a new color triplet and $SU(2)$ singlet quark with the same quantum number as the Standard Model down quark $d$. Thus, $\mathscr{D'}$ and $\mathscr{D}$ will mix and the actual SM down quark $d$ (and the actual mass eigenstate of the new heavy quark $d'$) will be a linear combination of $\mathscr{D'}$ and $\mathscr{D}$. On the other hand, the lepton content in this minimal Trinification model is given by:
\begin{eqnarray}
\begin{split}\label{Lepton}
	\psi_{E} &= \left(\begin{array}{ccc} \bar{\nu}'_L  & e'_L & e_L \\ \bar{e}'_L  & \nu'_L & \nu_L \\ e_R  & \nu_R & \nu' \end{array}\right) \sim  (1,3,3) \,,
\end{split}
\end{eqnarray}
where $\mathscr{L}=\left(e_L,\,\nu_L\right)$ is the usual lepton doublet while $\mathscr{E}=\left(e'_L,\,\nu'_L\right)$ denotes the heavier lepton doublet with the same hypercharge. Similar to the quark case, the two lepton doublets $\mathscr{L}$ and $\mathscr{E}$ will also mix and the actual mass eigenstates after mixing are denoted as $L$ and $E$ (shown more details later on).

In order to induce the breaking of $G_{\rm TR}$ to the SM gauge group, we introduce two scalar triplet fields $\Phi_1,\,\Phi_2$ which transform under the $G_{\rm TR}$ as $(1,3,3)$:
\begin{equation}
	\Phi_a = \left(\begin{array}{ccc} \phi_1^a  & \phi_2^a & \phi_3^a \\ S_1^a  & S_2^a & S_3^a\end{array}\right)\,,\qquad\left(a=1,2\right)\,,
\end{equation}
where $\phi_i^a,\,\left(i=1,2,3\right)$ denotes the Higgs doublets while $S_i^a,\,\left(i=1,2,3\right)$ denotes the singlets.
Note that it is argued in \cite{Pelaggi:2015kna} in order to match correctly three generations of the Standard Model matter content, three scalar triplets are required. However, in this work, for simplicity, we only focus on the case with two scalar triplets which are sufficient to address the correct flavour structure for the third generation.
The vacuum configuration of the scalar triplet is given as:
\begin{equation}
	\left\langle\Phi_1\right\rangle = \left(\begin{array}{ccc} u_1  & 0 & 0 \\ 0  & u_2 & 0\\0 & 0 & v_1\end{array}\right)\,,\qquad \left\langle\Phi_2\right\rangle = \left(\begin{array}{ccc} n_1  & 0 & n_3 \\ 0  & n_2 & 0\\v_2 & 0 & v_3\end{array}\right)\,,\label{scalar vacuum}
\end{equation}
where normally $v_3$ and $v_1$ play the role to break $G_{TR}$ to left right model i.e.~$SU(3)_C\times SU(3)_L\times SU(3)_R\rightarrow SU(3)_C\times SU(2)_R\times SU(2)_L\times U(1)_{B-L}$ while $v_2$ as an intermediate scale (between Trinification symmetry breaking scale and the electroweak scale) will further break the left-right symmetry to the Standard Model gauge group. On the other hand $u_1,\,u_2$ and $n_1,\,n_2,\,n_3$ are at the electroweak scale to trigger the electroweak symmetry breaking. In our work, for simplificity, we set $v_3=0$ ($v_1$ itself is sufficent to trigger the Trinification symmetry breaking) and assume $v_2$ is at the same scale as $v_1$ and thus, the Trinification model breaks directly to the SM without left-right model as an intermediate step.

\subsection{The Yukawa sector}
The Yukawa terms for the quark sector is given by:
\begin{eqnarray}
\mathcal{L}_{\rm Yuk}^Q & = & \overline{\psi}_{Q_L}\psi_{Q_R}\left(y_{\psi_{Q1}}\Phi_1+y_{\psi_{Q2}}\Phi_2\right)+\, \text{h.c.}\,\label{LYuk1}
\end{eqnarray}
In terms of the $SU(2)_L$ doublet, Eq.~(\ref{LYuk1}) reads:
\begin{equation}
\mathcal{L}_{\rm Yuk}^Q = m_{d'}\overline{d}'_L d'_R+\sum_{a=1}^2y_{\psi_{Qa}}\left[\overline{Q}\left(-s_\alpha d_R+c_\alpha d'_R\right)\phi_1^a+\overline{Q}u_R\phi_2^a+\overline{Q}\left(c_\alpha d_R+s_\alpha d'_R\right)\phi_3^a\right]+\, \text{h.c.}\,
\label{LYuk2}	                 
\end{equation}
where $Q$ denotes as the SM quark $SU(2)_L$ doublet, and $s_\alpha\equiv\sin\alpha,\,c_\alpha\equiv\cos\alpha$ are the mixing angles between $\mathscr{D},\,\mathscr{D'}$ defined below:
 \begin{equation}
	\left(\begin{array}{c} d \\ d' \end{array}\right)\;= \; \left(\begin{array}{cc}  -s_\alpha & c_\alpha \\ c_\alpha & s_\alpha\end{array} \right)\left(\begin{array}{c} \mathscr{D} \\ \mathscr{D'} \end{array}\right)\,,\qquad \tan\alpha=\frac{y_{\psi_{Q1}}v_1}{y_{\psi_{Q2}}v_2}\,,\label{mixing1}
\end{equation}
and the mass eigenvalue $m_{d'}$ is given by 
\begin{equation}
m_{d'}=\sqrt{y_{\psi_{Q1}}v_1^2+y_{\psi_{Q2}}v_2^2}\,.
\end{equation}

After electroweak symmetry breaking, $\phi_1^1,\,\phi_2^1$ obtain the VEV (vacuum expectation value) $u_1$ and $u_2$ respectively and the Standard Model quarks acquire masses (for simplicity, we have set $n_1=n_2=n_3=0$):
\begin{equation}
m_{t}=y_{\psi_{Q1}}u_2,\qquad\qquad m_{b}=y_{\psi_{Q1}}u_1s_{\alpha}\,,\label{top mass}
\end{equation}
where $m_t$ and $m_b$ denote the top and bottom quarks respectively. Since $(u_1^2+u_2^2)^{1/2}=246\,\rm{GeV}$ as the vacuum expectation value of the Higgs field at the electroweak scale, we thus have:
\begin{equation}
u_1=\frac{246}{\sqrt{1+\left(\frac{m_t}{m_b}\sin\alpha\right)^2}},\qquad u_2=\frac{246}{\sqrt{1+\left(\frac{m_b}{m_t}\frac{1}{\sin\alpha}\right)^2}}\,.
\end{equation}
Similarly, the Yukawa terms for the Lepton sector is given by:
\begin{eqnarray}
\mathcal{L}_{\rm Yuk}^{E} & = & \overline{\psi}_{E_L}\psi_{E_R}\left(y_{\psi_{E1}}\Phi_1+y_{\psi_{E2}}\Phi_2\right)+\, \text{h.c.}\,.\,\label{LYuk3}
\end{eqnarray}
In terms of the $SU(2)_L$ doublet, Eq.~(\ref{LYuk3}) can be written out explictly as:
\begin{equation}
\begin{split}
\mathcal{L}_{\rm Yuk}^E =& m_{E}\overline{E}_L E_R+\sum_{a=1}^2y_{\psi_{Ea}}\bigg\{-\left[\left(-c_\beta \nu_R-s_\beta\nu'\right)E_R-\left(c_\beta \overline{E}_L+s_\beta \overline{L}_L\right)e_R\right]\phi_1^a\\
+&\left(\overline{E}_L\nu'-\overline{L}_L\nu_R\right)\phi_2^a + \left[\left(s_\beta\nu_R-c_\beta\nu'\right)E_R-\left(-s_\beta\overline{E}+c_\beta\overline{L}\right)e_R\right]\phi_3^a\bigg\}\, +\text{h.c.}\,
\label{LYuk4}
\end{split}	                 
\end{equation}
where $L$ denotes as the SM quark $SU(2)_L$ doublet, and $s_\beta\equiv\sin\beta,\,c_\beta\equiv\cos\beta$ are the mixing angles between $\mathscr{L},\,\mathscr{E}$ defined below:
 \begin{equation}
	\left(\begin{array}{c} E \\ L \end{array}\right)\;= \; \left(\begin{array}{cc}  -s_\beta & c_\beta \\ c_\beta & s_\beta\end{array} \right)\left(\begin{array}{c} \mathscr{E} \\ \mathscr{L} \end{array}\right)\,,\qquad \tan\beta=\frac{y_{\psi_{E1}}v_1}{y_{\psi_{E2}}v_2}\,,\label{mixing2}
\end{equation}
and the mass eigenvalue $m_{E}$ is given by 
\begin{equation}
m_{E}=\sqrt{y_{\psi_{E1}}v_1^2+y_{\psi_{E2}}v_2^2}\,.
\end{equation}
After electroweak symmetry breaking, the SM lepton masses are given by:
\begin{equation}
m_e=y_{\psi_{E1}}u_1s_\beta\,,\qquad m_{\nu_L,\nu_R}=y_{\psi_{E1}}u_2\,,\qquad m_{\nu'}\sim\frac{y_{\psi_{E1}}^2u_1u_2s_\beta}{m_E}\,.\label{neutrino_mass_tree}
\end{equation}
It is clear that the above tree level neutrino mass Eq.~\eqref{neutrino_mass_tree} has the problem that the left and right handed neutrino masses are degenerate. Fortunately, the radiative loop corrections can cure this problem and provides reasonable neutrino mass spectrum.

\subsection{Radiative Neutrino Mass}
\begin{figure}[t]
\centering
\includegraphics[width=0.6\columnwidth]{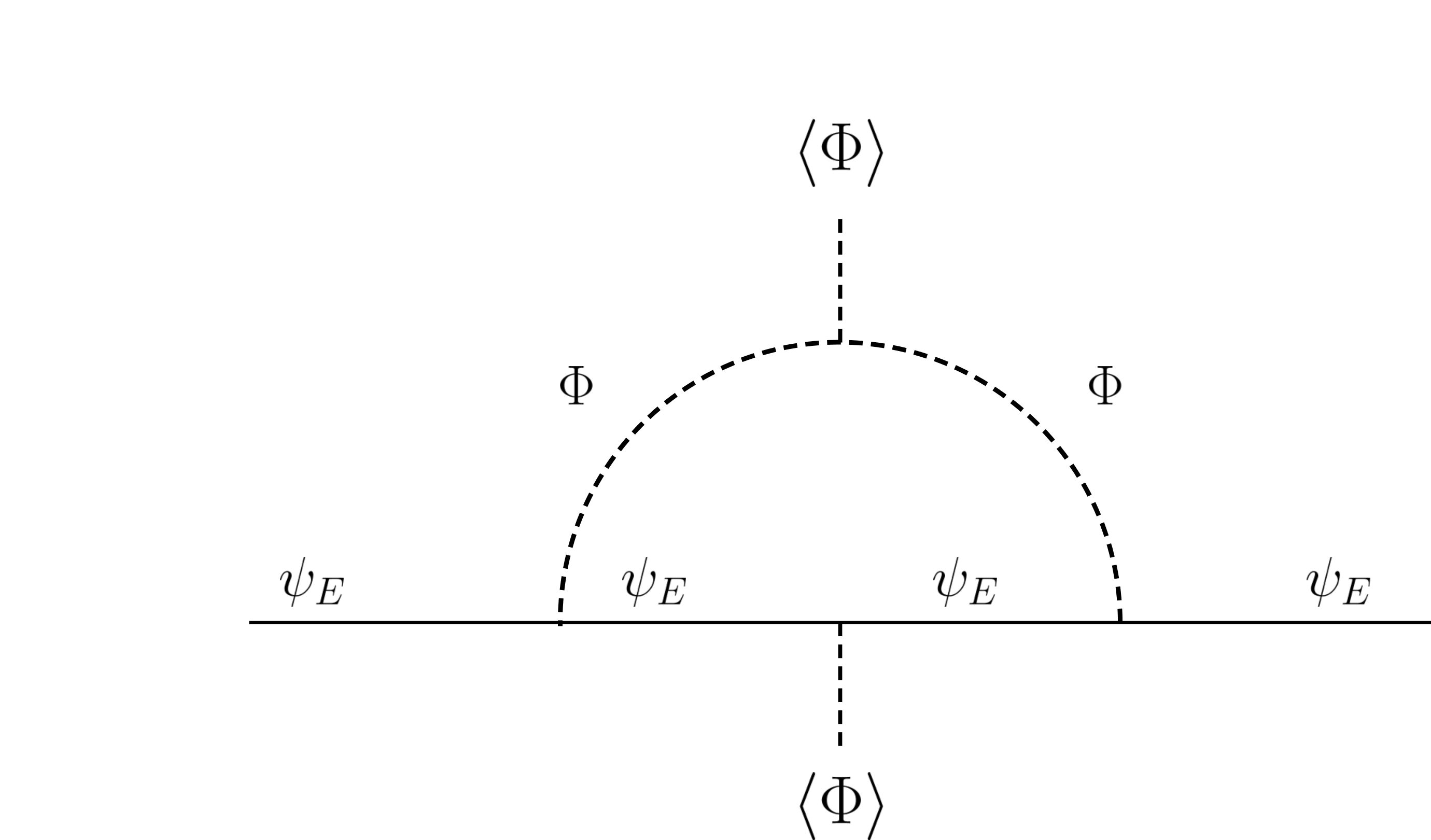}\hspace{0.05\columnwidth}
\caption{This figure shows the one loop radiative correction to the neutrino mass.}
\label{loop neutrino}
\end{figure}

From fig.~\ref{loop neutrino}, the neutrino masses receive large radiative loop contributions. The total contribution to the two point function is proportional to:
\begin{equation}
F_E=\frac{m_E}{\left(4\pi\right)^2}\frac{1}{2}\left(\frac{m_{H1}^2}{m_E^2-m_{H1}^2}\log\frac{m_{H1}^2}{m_E^2}-\frac{m_{H2}^2}{m_E^2-m_{H2}^2}\log\frac{m_{H2}^2}{m_E^2}\right)\,,\label{neutrino two point functions}
\end{equation}
where $m_{H1}$ and $m_{H2}$ are the masses of the two Higgs doublet fields. 

The one-loop neutrino mass matrix is thus written as:
\begin{equation}
M_{\nu}^{\rm{1loop}}=\left[
\begin{array}{ccc}
 0 & -y_{\psi_{E1}}u_2 & 0\\
 -y_{\psi_{E1}}u_2 & s_{\alpha-\beta}c_\beta y_{\psi_{E1}}^2F_E & \left(s_{2\beta}s_\alpha-c_\alpha\right)y_{\psi_{E1}}^2F_E \\
 0 & \left(s_{2\beta}s_\alpha-c_\alpha\right)y_{\psi_{E1}}^2F_E & c_{\alpha-\beta}s_\beta y_{\psi_{E1}}^2 F_E
\end{array}
\right]\,.\label{neutrino mass 1loop}
\end{equation}
To obtain a phenomenological viable neutrino mass, the elements of the matrix Eq.~\eqref{neutrino mass 1loop} should have the following form:
\begin{equation}
M_{\nu}^{\rm{1loop}}=\left[
\begin{array}{ccc}
 0 & 10\,\rm{GeV} & 0\\
10\,\rm{GeV} & 0\--1\,\rm{TeV} & 0.33\--1\,\rm{TeV} \\
 0 & 0.33\--1\,\rm{TeV} & 1\,\rm{KeV}
\end{array}
\right]\,,\label{pheno neutrino mass}
\end{equation}
where the the three mass eigenvalues will correspond to the physical mass of the two sterile neutrinos ($\nu_R,\,\nu'$) and the SM neutrino $\nu_L$.
Comparing Eq.~\eqref{neutrino mass 1loop} and Eq.~\eqref{pheno neutrino mass}, we find from the $\left(3,\,3\right)$ element of the matrix that $c_{\alpha-\beta}s_\beta\sim 10^{-8}\--10^{-9}$. There are only two solutions to satisfy the above constraint either $\beta<<1$ or $\alpha-\beta-\frac{\pi}{2}\sim\pm10^{-9}$. For $\beta<<1$, from Eq.~\eqref{neutrino_mass_tree}, it will lead to extremely large Yukawa $y_{\psi_{E1}}$ which is not acceptable. Thus, we choose:
\begin{equation} 
\alpha-\beta-\frac{\pi}{2}\sim\pm10^{-9}\label{angle relationship}
\end{equation}
in the later on analysis.

\subsection{Colour Scalar Field and Proton Decay Issue}
Sometimes, it occurs in the references that colour scalars are introduced to make the Yukawa sector satisfy cyclic permutation symmetry. The introduction of the colour scalar fields has the advantage to incorporate much larger radiative corrections to the neutrinos. In this case, in Eq.~\eqref{neutrino two point functions}, the $m_{H1}$ should be replaced by the new coloured scalar field $m_{B_{H1}}$ and the lepton doublet $m_E$ should be replaced by the heavy quark $m_B$. The radiative corrections to the neutrino mass can be much larger because colour scalar fields are not constrained to be around electroweak scale and thus $m_{B_{H1}}>>m_{H1}$ resulting a much larger radiative loop corrections.

However, coloured scalar fields have the problem to generate tree level proton decay and thus in this work we only introduce the scalar fields which are colour singlet. The leptonic scalar in this work do not introduce proton decay because it carries the same baryon number in both the quark and leptonic Yukawa interactions.

\subsection{The Scalar sector}
In this minimal model, we have only introduced two Higgs triplet $\Phi_1$ and $\Phi_2$. The most general scalar potentials can be written as:
\begin{equation}
V=V_{1111}+V_{2222}+V_{1122}+V_{1222}+V_{1112}\,,
\end{equation}
where in the following we could further simply the potential by introducing a $Z_2$ discrete symmetry between $\Phi_1$ and $\Phi_2$ and thus only the terms $V_{1111},\,V_{2222},\,V_{1122}$ remain. We have:
\begin{equation}
\begin{split}
V_{1111}&=\lambda_{1a}\trace\left(\Phi_1^\dagger\Phi_1\right)^2+\lambda_{1b}\trace\left(\Phi_1^\dagger\Phi_1\Phi_1^\dagger\Phi_1\right)\\
V_{2222}&=\lambda_{2a}\trace\left(\Phi_2^\dagger\Phi_2\right)^2+\lambda_{2b}\trace\left(\Phi_2^\dagger\Phi_2\Phi_2^\dagger\Phi_2\right)\\
V_{1122}&=\lambda_a\trace\left(\Phi_1^\dagger\Phi_1\right)\trace\left(\Phi_2^\dagger\Phi_2\right)+\lambda_b\left\vert\trace\left(\Phi_1^\dagger\Phi_2\right)\right\vert^2\\
&+\lambda_c\trace\left(\Phi_1^\dagger\Phi_1\Phi_2^\dagger\Phi_2\right)+\lambda_d\trace\left(\Phi_1\Phi_1^\dagger\Phi_2\Phi_2^\dagger\right)\\
&+\rm{Re}\lambda_e\trace\left(\Phi_1^\dagger\Phi_2\right)^2+\rm{Re}\lambda_f\trace\left(\Phi_1^\dagger\Phi_2\Phi_1^\dagger\Phi_2\right)\,.\label{triplet potential}
\end{split}
\end{equation}

\section{Renormalization group analysis}\label{Sec3}

 In this section, we perform a RG analysis of the minimal Trinification model. After introducing the Large-$N$ beta functions of the gauge, Yukawa and quartic couplings, we will demonstrate the existence of UV fixed point solutions. By using the RG equation as a bridge to connect UV and IR, we will then study the IR phenomenological implications. All  gauge, Yukawa and scalar couplings are listed in table~\ref{couplings}, while their corresponding one loop RG beta functions are listed in appendix~\ref{RG_all}.
\begin{table}[t!]
\centering
  \begin{tabular}{|| l | l | l ||}
    	\hline
Gauge Couplings & Yukawa Couplings & Scalar Couplings \\ \hline
$SU(3)_C:\,g_c$ & $\psi_{Q_{L/R}}: y_{\psi_{Q1}},\,y_{\psi_{Q2}}$ & $V_{1111}: \lambda_{1a},\,\lambda_{1b}$\\\hline
$SU(3)_L:\,g_L$ & $\psi_{E}: y_{\psi_{E1}},\,y_{\psi_{E2}}$ & $V_{2222}: \lambda_{2a},\,\lambda_{2b}$\\ \hline
$SU(3)_R:\,g_R$ & & $V_{1122}: \lambda_a,\,\lambda_b,\,\lambda_c,\,\lambda_d,\,\lambda_e,\,\lambda_f$\\ \hline
\end{tabular}
\caption{\small Gauge, Yukawa and scalar quartic couplings of the Trinification model.}
\label{couplings}
\end{table}

\mathversion{bold}
\subsection{Large-$N$ beta function}
\mathversion{normal}

\begin{figure}[t]
\centering
\subfigure[]{
\label{bubble_1}
\begin{minipage}{6cm}
\centering
\includegraphics[width=1\columnwidth]{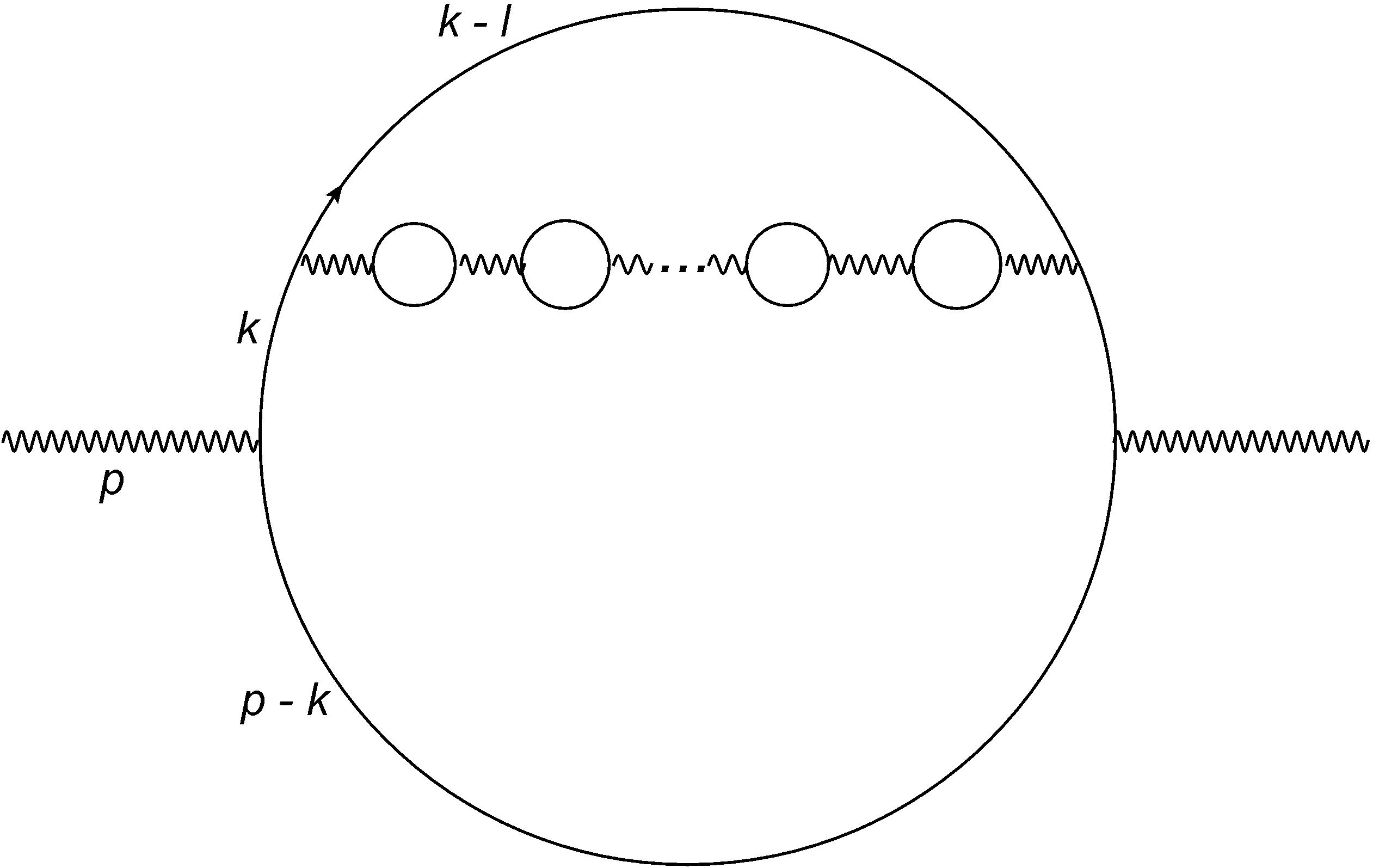}\hspace{0.05\columnwidth}
\end{minipage}
}
\subfigure[]{
\label{bubble_2}
\begin{minipage}{6cm}
\centering
\includegraphics[width=1\columnwidth]{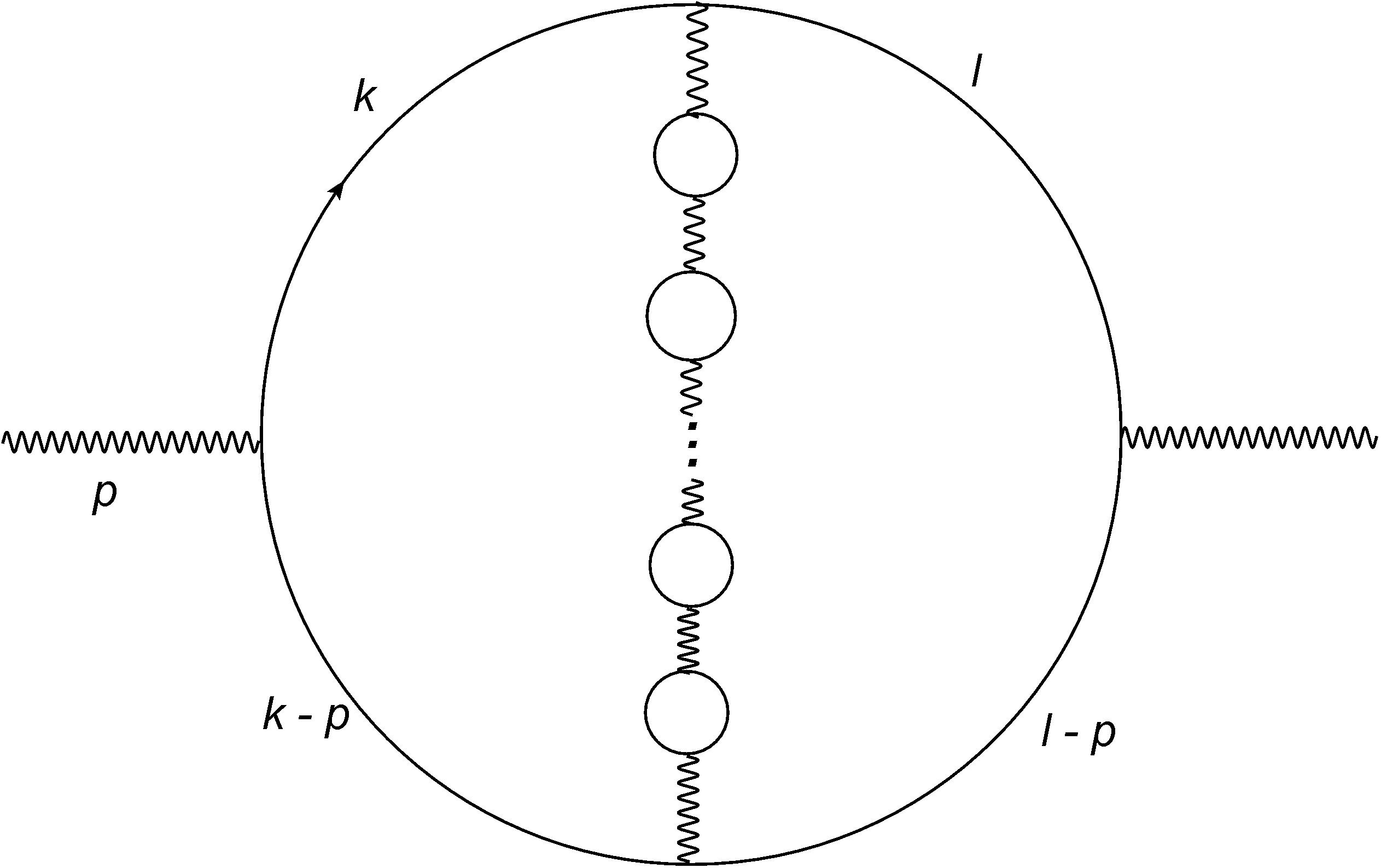}\hspace{0.05\columnwidth}
\end{minipage}
}
\subfigure[]{
\label{bubble_3}
\begin{minipage}{6cm}
\centering
\includegraphics[width=1\columnwidth]{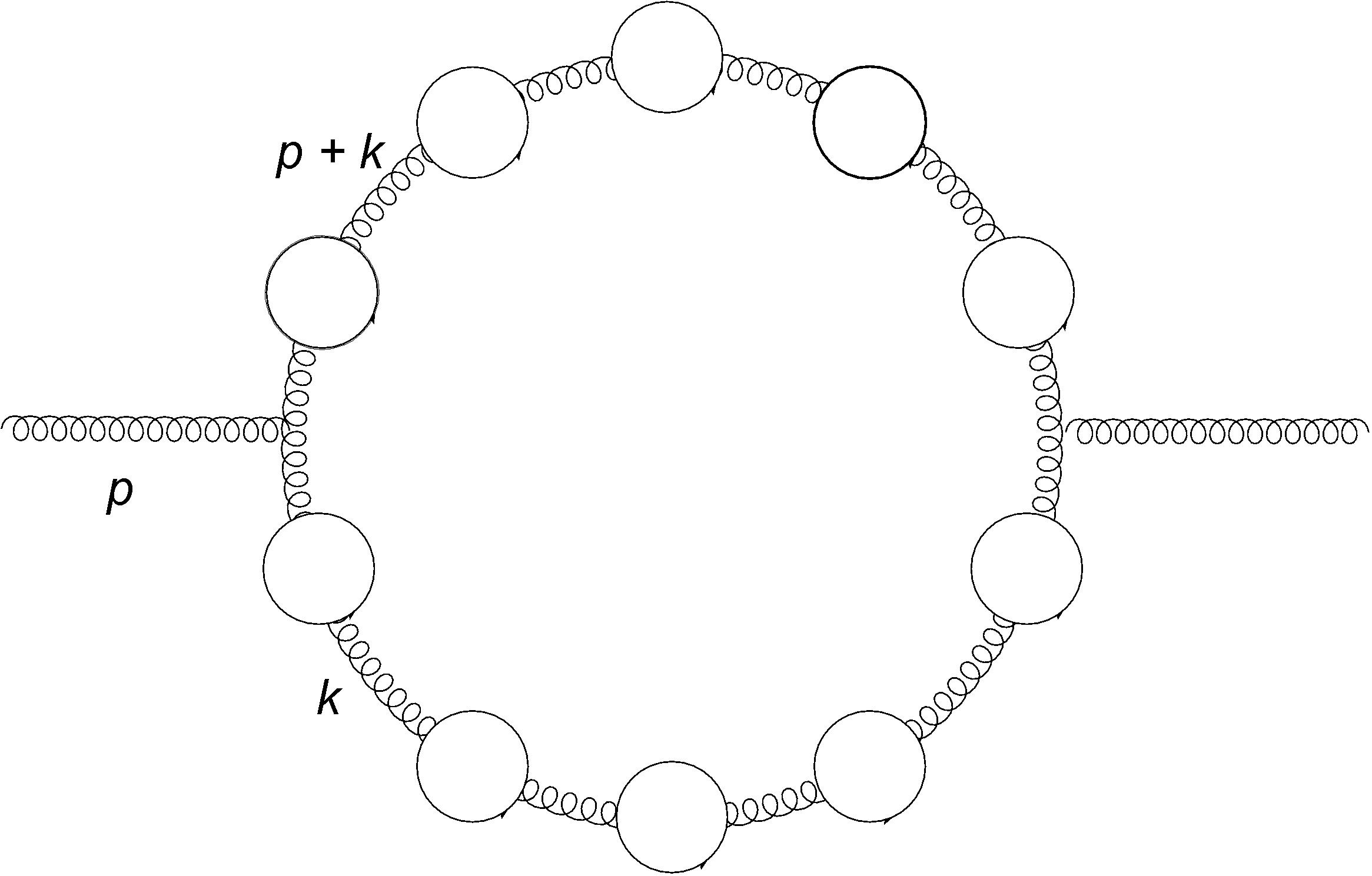}\hspace{0.05\columnwidth}
\end{minipage}
}
\subfigure[]{
\label{gauge_cancel}
\begin{minipage}{6cm}
\centering
\includegraphics[width=1\columnwidth]{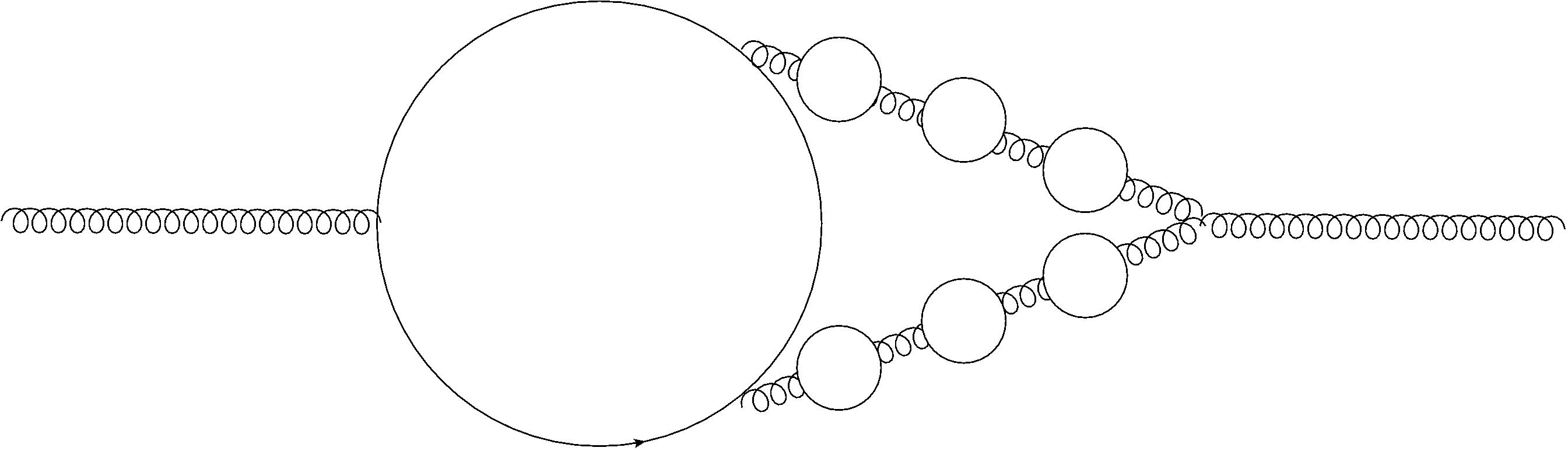}\hspace{0.05\columnwidth}
\end{minipage}
}
\caption{Feynman diagrams for gauge field renormalization at order $1/N_F$. Diagrams (a) and (b) are present in both the Abelian and non-Abelian 2-point functions, while (c) and (d) only exist in the non-Abelian theory. }
\label{bubble diagram}
\end{figure}

To proceed, we work in the large $N_F$ limit, employing the $1/N_F$ expansion~\cite{PalanquesMestre:1983zy,Gracey:1996he,Holdom:2010qs,Pica:2010mt},  which was recently first applied to the whole SM~\cite{Mann:2017wzh} and to the minimal Pati-Salam model~\cite{Molinaro:2018kjz}.  We introduce $N_F\gg 1$ vector-like fermions, which transform non-trivially under $G_{\rm{TR}}$. In this framework, the leading order RG contributions in the $1/N_F$ expansion of the relevant Feynman diagrams can be resummed, as shown 
in fig.~\ref{bubble diagram} (only gauge coupling cases are shown) and a closed form of the resummation is provided. This non-perturbative effect induces an interacting fixed point for both the Abelian and non-Abelian gauge interactions of the SM \cite{Mann:2017wzh}, whose existence is guaranteed due to the pole structure  in the closed form expression~\cite{Gracey:1996he, Holdom:2010qs}.

We introduce three sets of vector-like fermions charged under $G_{\rm TR}$
with the following charge assignment:
\begin{equation}
N_{F_{C}}\left(3,1,1\right)\oplus N_{F_{L}}\left(1,3,1\right)\oplus N_{F_{R}}\left(1,1,3\right)\,,
\end{equation}
where we have chosen each set of vector-like fermions to have non-trivial charges only under one simple gauge group to avoid  extra contributions in the summation of the semi-simple group.

\mathversion{bold}
\subsection{Large-$N$ gauge beta function}
\mathversion{normal}

To  leading order in $1/N_F$, the higher order (ho) contributions of the bubble diagrams  (i.e.~those in fig.~\ref{bubble diagram}) to the gauge beta functions were first calculated in \cite{PalanquesMestre:1983zy} for only an Abelian group and later generalized to the  non-abelian simple group case \cite{Holdom:2010qs}. Here we summarize the results. 

The leading $1/N_F$ order bubble diagrams' contributions are given by:
\begin{equation}
\beta^{{\rm ho}}_{ i}=\frac{2A_i\alpha_i}{3}\frac{H_{1_i}(A_i)}{N_{F_i}},\quad\alpha_i\equiv\frac{g_i^2}{\left(4\pi\right)^2}~~\left(i=L,\,R,\,C\right)\,,\label{higher order contribution}
\end{equation}
with the functions $H_{1i}$ and the t'Hooft couplings $A_i$ given by
\begin{equation}
\begin{split}
A_i&=4\alpha_iT_RN_{F_i}\\
H_{1_i}&=-\frac{11}{2}N_{c_i}+\int_0^{A_i/3}I_1(x)I_2(x)dx\qquad\left(N_{c_i}=3\right)\\
I_1(x)&=\frac{\left(1+x\right)\left(2x-1\right)^2\left(2x-3\right)^2\sin\left(\pi x\right)^3}{\left(x-2\right)\pi^3}\times\left(\Gamma\left(x-1\right)^2\Gamma\left(-2x\right)\right)\\
I_2(x)&=\frac{N_{c_i}^2-1}{N_{c_i}}+\frac{\left(20-43x+32x^2-14x^3+4x^4\right)}{2\left(2x-1\right)\left(2x-3\right)\left(1-x^2\right)}N_{c_i}\,.
\end{split}
\end{equation}
The Dynkin indices are $T_R=1/2~(N_{c_i})$ for the fundamental (adjoint) representation.
The RG functions of the gauge couplings (see Appendix \ref{RG_all}), including the contributions from the resummed bubble diagrams, are 
\begin{equation}
\begin{split}
\beta_{\alpha_{L}}^{tot}&=\beta_{\alpha_{L}}^{1loop}+\beta_{\alpha_{L}}^{\rm{ho}}=-\left(10+n_H\right)\alpha_{L}^2+\frac{2A_{L}\alpha_{L}}{3}\frac{H_{1_{L}}\left(A_{L}\right)}{N_{F_{L}}}\\
\beta_{\alpha_{R}}^{tot}&=\beta_{\alpha_{R}}^{1loop}+\beta_{\alpha_{R}}^{\rm{ho}}=-\left(10+n_H\right)\alpha_{R}^2+\frac{2A_{R}\alpha_{R}}{3}\frac{H_{1_{R}}\left(A_{R}\right)}{N_{F_{R}}}\\
\beta_{\alpha_{c}}^{tot}&=\beta_{\alpha_{c}}^{1loop}+\beta_{\alpha_{c}}^{\rm{ho}}=-10\alpha_{c}^2+\frac{2A_{c}\alpha_{c}}{3}\frac{H_{1_c}\left(A_{c}\right)}{N_{F_{C}}}\,,\\
\end{split}
\label{Gauge couplings RG Bubble}
\end{equation}
where $n_H$ denotes the number of scalar triplets ($n_H=2$ in our case). The $\beta_{\alpha_{L}}^{1loop}$, $\beta_{\alpha_{R}}^{1loop}$, $\beta_{\alpha_{c}}^{1loop}$   denote   the original one loop RG beta functions of the three gauge couplings without bubble diagram contributions, whereas $\beta_{\alpha_{L}}^{tot},\,\beta_{\alpha_{R}}^{tot},\,\beta_{\alpha_{c}}^{tot}$ are the total RG beta functions including the higher order bubble diagram contributions up to $1/N_F$ order. The reason that only one loop RG beta functions of the gauge couplings are used will be clear later on.

Hence the UV fixed point for the gauge coupling sub-system $\left(g_c,\,g_L,\,g_R\right)$ is guaranteed by the pole structure in the bubble diagram summation. 
For all the non-abelian gauge groups, the poles in the functions $H_{1_i} \left(i=L,\,R,\,C\right)$ always occur at $A_i=3$, which determines the UV fixed point of the non-abelian gauge couplings when the set of $N_{F_i}$ is chosen. The IR initial conditions of $g_L$, $g_R$ and $g_c$ are obtained by using the matching conditions of Eq.~\eqref{matching_1} and the SM couplings are running from the EW scale to the Trinification symmetry breaking scale. Also, for simplicity, we have assumed all the vector-like fermions were introduced at the Trinification symmetry breaking scale $v_{T}$.  This is in contrast to the Pati-Salam model whose symmetry breaking scale is severely constrained by the kaon decay process $K_L\rightarrow\mu^{\pm}e^{\mp}$ (see e.g.~\cite{Volkas:1995yn,Valencia:1994cj}), which implies the symmetry breaking scale must be larger than $2000\,\rm{TeV}$.  Trinification symmetry breaking is only constrained by the masses of the extra gauge bosons such as the $Z'$ and $W_{R}$ which have lower bounds of a few TeV.  Trinification therefore has the advantage of being within 
experimental reach of an upgraded LHC and future colliders.


\mathversion{bold}
\subsection{Large-$N$ Yukawa and quartic beta function}
\mathversion{normal}

 In the previous section, we exhibited the bubble diagram contributions in only the gauge coupling subsystem and  presented the large $N$ gauge beta functions.  We now consider  bubble diagram contributions to the Yukawa and quartic beta functions~\cite{Antipin:2018zdg,Kowalska:2017pkt}. In the following we briefly summarize  the procedure.

If the beta functions of quartic and Yukawa couplings are already known to 1-loop order, the corresponding large $N_F$ beta functions (at leading $1/N_F$ order) can be obtained by simply employing the following recipe.
The large-$N_F$ Yukawa beta function can be written in the following compact form
\begin{equation}
\begin{split}
\beta_y&=c_1y^3 + y \sum_{\alpha}c_\alpha g^2_\alpha I_y\left(A_\alpha\right),\quad\rm{with}\\\label{eq-simplifiedyukawa}
I_y\left(A_\alpha\right)&=H_\phi\left(0,\tfrac{2}{3}A_\alpha\right)\left(1+A_\alpha\frac{C_2\left(R_\phi^\alpha\right)}{6\left(C_2\left(R_{\chi}^\alpha\right)+C_2\left(R_{\xi}^\alpha\right)\right)}\right)\\
H_\phi(x) &=H_0(x)= \dfrac{(1 - \tfrac{x}{3}) \Gamma(4-x)}{3 \Gamma^2(2 - \tfrac{x}{2}) \Gamma(3 - \tfrac{x}{2}) \Gamma(1 + \tfrac{x}{2})}\,,
\end{split}
\end{equation}
 where the information of the resumed fermion bubbles is already encoded and $c_1,\,c_\alpha$ are the standard 1-loop coefficients for the Yukawa beta function while $C_2(R_\phi^\alpha),\,C_2(R_{\chi}^\alpha),\,C_2(R_{\xi}^\alpha)$ are the Casimir operators of the corresponding scalar and fermion fields. Thus, when $c_1$ and $c_\alpha$ are known, the full Yukawa beta function including the bubble diagram contributions can be obtained.
Similarly, for the quartic coupling we write
\begin{equation}
\beta_\lambda=c_1\lambda^2+\lambda \sum_{\alpha}c_\alpha \,g^2_\alpha\,I_{\lambda g^2}\left(A_\alpha\right)+\sum_{\alpha} c'_\alpha \,g_\alpha^4\,I_{g^4} \left(A_\alpha\right) +\sum_{\alpha < \beta}c_{\alpha\beta}\, g_\alpha^2 g_\beta^2 \,I^{tot}_{g_1^2g_2^2}\left(A_\alpha,\,A_\beta\right)\,,\label{quartic_bubble_beta}
\end{equation}
with $c_1,\,c_\alpha,\,c'_\alpha,\,c_{\alpha\beta}$  the known 1-loop coefficients
for the quartic beta function and 
\begin{equation}
\begin{split}
I_{\lambda g^2}\left(A_\alpha\right) &=H_\phi\left(0,\tfrac{2}{3}A_\alpha\right)\\
I_{g^4}\left(A_\alpha\right)&=H_\lambda\left(1,\tfrac{2}{3}A_\alpha\right)+A_\alpha\frac{dH_\lambda\left(1,\tfrac{2}{3}A_\alpha\right)}{dA_\alpha}\\
I_{g_1^2g_2^2}^{tot}\left(A_\alpha,\,A_\beta\right)&=\frac{1}{3}\left[I_{g_1^2g_2^2}\left(A_\alpha,\,0\right)+I_{g_1^2g_2^2}\left(0,\,A_\beta\right)+I_{g_1^2g_2^2}\left(A_\alpha,\,A_\beta\right)\right]\\
I_{g_1^2g_2^2}\left(A_\alpha,\,A_\beta\right)&=\frac{1}{A_\alpha-A_\beta}\left[A_\alpha H_\lambda\left(1,\tfrac{2}{3}A_\alpha\right)-A_\beta H_\lambda\left(1,\tfrac{2}{3}A_\beta\right)\right],\quad\rm{where}\\
H_\lambda(1,x) &= (1-\tfrac{x}{4}) H_0(x)=\dfrac{ (1-\tfrac{x}{4})(1 - \tfrac{x}{3}) \Gamma(4-x)}{3 \Gamma^2(2 - \tfrac{x}{2}) \Gamma(3 - \tfrac{x}{2}) \Gamma(1 + \tfrac{x}{2})} 
\end{split}
\end{equation} 
are from the resumed fermion bubbles. 
Thus we have now  the full quartic beta function including the bubble diagram contributions when $c_1,\,c_\alpha,\,c_\alpha',\,c_{\alpha\beta}$ are known. Following the above recipe, the bubble diagram improved Yukawa beta function $\beta_{y_{Q1}}$, for example, can be written as
\begin{equation}
\begin{split}
(4 \pi )^2\beta_{y_{Q1}}&=\left(-4 g_L^2 I_y\left(A_L\right)-4g_R^2 I_y\left(A_R\right)- 8 g_c^2 I_y\left(A_3\right)+6 y_{\psi_{Q1}}^2+ 6y_{\psi_{Q2}}^2+2y_{\psi_{E1}}^2\right)y_{\psi_{Q1}}\\
&+12 y_{\psi_{E1}}y_{\psi_{E2}}y_{\psi_{Q2}}\,.
\end{split}
\end{equation}
The bubble diagram improved quartic beta function $\beta_{\lambda_{1a}}$ reads
\begin{equation}
\begin{split}
(4 \pi )^2\beta_{\lambda_{1a}}&=52 \lambda _{1a}^2+12\lambda_{1b}^2+2\lambda_a\lambda_b+6\lambda_a\lambda_c+6\lambda_a\lambda_d+2\lambda_c\lambda_d+9\lambda_a^2+\lambda_b^2+4\lambda_e^2+4\lambda_f^2\\
&+\lambda _{1a} \left(-16 g_L^2 I_{\lambda g^2}\left(A_L\right)-16 g_R^2 I_{\lambda g^2}\left(A_R\right)+48 \lambda _{1b}+8 y_{\psi_{E1}}^2+12 y_{\psi_{Q1}}^2\right)\\
&+\frac{10}{3} g_L^2 g_R^2 \times\frac{1}{3}\left(I_{g_1^2g_2^2}\left(A_L,A_R\right)+I_{g_1^2g_2^2}\left(0,A_R\right)+I_{g_1^2g_2^2}\left(A_L,0\right)\right)\\
&+\frac{11}{12} g_L^4 I_{g^4}\left(A_L\right)+\frac{11}{12} g_R^4 I_{g^4}\left(A_R\right)
-2 y_{\psi_{E1}}^4\,.
\end{split}
\end{equation}

\subsection{Symmetric and Asymmetric Cases}

From the Yukawa beta functions in Appendix \ref{App1}, we notice that there exists a symmetry between $y_{\psi_{Q1}}$ and $y_{\psi_{Q2}}$ as well as $y_{\psi_{E1}}$ and $y_{\psi_{E2}}$. 
Consequently there is not a 
unique  UV fixed point solution for the Yukawa couplings when the $N_{F}$ is fixed (actually there could be infinitely many solutions).

It is convenient to divide the fixed point solutions into two cases: symmetric and asymmetric. 
By `symmetric' we mean that $y_{\psi_{Q1}}=y_{\psi_{Q2}}$ and $y_{\psi_{E1}}=y_{\psi_{E2}}$ at the UV fixed point. In the symmetric case, all the coupling values at the UV fixed point can be determined when $N_F$ is fixed. In addition, from Eq.~\eqref{mixing1} and Eq.~\eqref{mixing2}, we know $\left\vert\tan\alpha\right\vert=\left\vert\tan\beta\right\vert=\frac{v_1}{v_2}$. Furthermore, a phenomenologically viable neutrino mass requires $\left\vert\alpha-\beta\right\vert\sim\frac{\pi}{2}$, providing $\left\vert\tan\alpha\right\vert=\left\vert\tan\beta\right\vert^{-1}$. Hence
\begin{equation}
\left\vert\tan\alpha\right\vert=\left\vert\tan\beta\right\vert=\frac{v_1}{v_2}=1\,.
\end{equation}

By `asymmetric' we mean that these Yukawa couplings are no longer constrained to be equal at the UV fixed point;  the dynamical constraints from requiring a UV interacting fixed point in this case are not sufficient to provide a unique set of UV fixed point solutions. We therefore add one more phenomenological constraint:
\begin{equation}
\frac{y_b}{y_\tau}=\frac{y_{\psi_{Q1}}}{y_{\psi_{E1}}}\frac{\sin\alpha}{\sin\beta}\sim2 \label{ratio1}
\end{equation}
which is that of requiring the ratio of bottom quark mass and the tau lepton mass to be around 2.  
We can further get rid of the $v_1^2/v_2^2$ dependence in \eqref{ratio1} and write it as a function of $\left(y_{\psi_{Q1}},\,y_{\psi_{Q2}}\,y_{\psi_{E1}},\,y_{\psi_{E2}}\right)$ only. By using $\left\vert\tan\alpha\right\vert=\left\vert\tan\beta\right\vert^{-1}$ we obtain
\begin{equation}
\frac{v_1^2}{v_2^2}=\frac{y_{\psi_{Q2}}y_{\psi_{E2}}}{y_{\psi_{Q1}}y_{\psi_{E1}}}\,,
\end{equation}
yielding
\begin{equation}
\frac{y_b}{y_\tau}=\frac{y_{\psi_{Q1}}^{3/2}}{y_{\psi_{E1}}^{3/2}}\frac{\left(y_{\psi_{Q2}}y_{\psi_{E2}}y_{\psi_{E1}}+y_{\psi_{E2}}^2y_{\psi_{Q1}}\right)^{1/2}}{\left(y_{\psi_{Q1}}y_{\psi_{Q2}}y_{\psi_{E2}}+y_{\psi_{Q2}}^2y_{\psi_{E1}}\right)^{1/2}}\sim2 \label{ratio2}
\end{equation}
and using~\eqref{ratio2}, we are able find  UV fixed point solutions.

\subsection{UV fixed point solutions in the gauge-Yukawa-quartic system}

 In this section we will prove the existence of  UV fixed point solutions for the whole gauge-Yukawa-quartic system given in table~\ref{couplings}. Note that the gauge couplings at the UV fixed point can be treated as background values (i.e.~as constants in the RG functions of other couplings) since their values at the UV fixed point are fixed when $N_F$ is chosen.
By using the one loop RG functions in appendix \ref{RG_all} and following the recipes from~\eqref{eq-simplifiedyukawa} and~\eqref{quartic_bubble_beta}, we obtain the total large-$N$ beta functions of all the Yukawa and scalar couplings. To find the UV fixed point solutions, we set $\{\beta_i=0\}$,  where $i$ denotes all the Yukawa and scalar couplings in table~\ref{couplings}.

\begin{table}[t!]
\centering
  \begin{tabular}{|| l | l | l | l | l | l | l | l | l | l | l | l ||}
    	\hline
  $\lambda_{1a}$ & $\lambda_{1b}$ & $\lambda_a$ & $\lambda_b$ & $\lambda_c$ & $\lambda_d$ & $\lambda_e$ & $\lambda_f$ & $y_{\psi_{Q1}}$ & $y_{\psi_{Q2}}$ & $y_{\psi_{E1}}$ & $y_{\psi_{E2}}$ \\ \hline
   0.10 & -0.03 & -0.82 & 0.43 & 1.22 &  -0.27 & -0.30 & 0.50 & 0.80 & 0.80 & 0.24 & 0.24 \\ \hline
  Irev & Irev  & Irev  & Irev  & Irev  &  Rev  & Irev  & Irev  & Irev  & Irev  & Irev  & Irev  \\ \hline
\end{tabular}
\caption{\small This table summarizes the sample UV fixed point solution in the symmetric case with sample value ($N_{F_C}=43,\,N_{F_L}=93,\,N_{F_R}=182$)
 involving the bubble diagram contributions in the Yukawa and quartic RG beta functions. Classifications of the UV fixed point solutions of the couplings with relevant (Rev) and irrelevant (Irev) characteristics are listed. }
\label{UV fixed point_1}
\end{table}
The analysis provides quite a few UV candidate fixed points for different choices of $N_F$. For example, in the symmetric case, for $N_{F_C}=43,\,N_{F_L}=93,\,N_{F_R}=182$, we find 56 sets of candidate UV fixed point solutions.  Requiring the vacuum stability conditions~\cite{Holthausen:2009uc} 
\begin{equation}
\lambda_{1a}+\lambda_{1b}>0\quad\left(\rm{for}\,\lambda_{1b}<0\right);\qquad\lambda_{1a}+\frac{1}{3}\lambda_{1b}>0\quad\left(\rm{for}\,\lambda_{1b}>0\right)\label{vacuum_stability}\,,
\end{equation}
which guarantee that the scalar potential is bounded from below, 
reduces  the number of candidate   UV fixed point solutions to 18.
One of these solutions in the symmetric case is shown in table~\ref{UV fixed point_1}, where we have also classified the fixed point solutions according to relevant (Rev) and irrelevant (Irev) characteristics. By choosing the RG flow direction from UV to IR, relevant and irrelevant correspond respectively to the RG flows running away from UV fixed point or towards UV fixed point.
From table~\ref{UV fixed point_1}, it is clear that most of the couplings are UV irrelevant;  only $\lambda_d$ is relevant,  implying the system is highly predictive. An alternative candidate solution in the asymmetric case is given in tables~\ref{UV fixed point_2} and~\ref{UV fixed point_3}, where we see that the symmetry between $\left(y_{\psi_{Q1}},\,y_{\psi_{E1}}\right)$ and $\left(y_{\psi_{Q2}},\,y_{\psi_{E2}}\right)$ is broken.
\begin{table}
\centering
  \begin{tabular}{|| l | l | l | l ||}
    	\hline
  $y_{\psi_{Q1}}$ & $y_{\psi_{Q2}}$ & $y_{\psi_{E1}}$ & $y_{\psi_{E2}}$ \\ \hline
    0.78 & 0.56 & 0.42  & 0.30\\ \hline
  Irev & Irev  & Irev  & Irev  \\ \hline
\end{tabular}
\caption{\small This table summarizes the sample UV fixed point solution (for Yukawa couplings) in the asymmetric case with sample value ($N_{F_C}=95,\,N_{F_L}=165,\,N_{F_R}=62$)
 involving the bubble diagram contributions in the Yukawa and quartic RG beta functions. Classifications of the UV fixed point solutions of the couplings with relevant (Rev) and irrelevant (Irev) characteristics are listed. }
\label{UV fixed point_2}
\end{table}

\begin{table}
\centering
  \begin{tabular}{|| l | l | l | l | l | l | l | l | l | l ||}
    	\hline
  $\lambda_{1a}$ & $\lambda_{1b}$ & $\lambda_{2a}$ & $\lambda_{2b}$ & $\lambda_a$ & $\lambda_b$ & $\lambda_c$ & $\lambda_d$ & $\lambda_e$ & $\lambda_f$ \\ \hline
   -0.08 & 0.26 & 0.02 & 0.02 & -0.90 & 0.14 & 0.12 & 1.11 & -0.36  & 0.48\\ \hline
  Irev & Irev  & Rev  & Rev  & Rev  &  Irev  & Rev  & Irev & Irev & Irev \\ \hline
\end{tabular}
\caption{\small This table summarizes the sample UV fixed point solution (for quartic couplings) in the asymmetric case with sample value ($N_{F_C}=95,\,N_{F_L}=165,\,N_{F_R}=62$)
 involving the bubble diagram contributions in the Yukawa and quartic RG beta functions. Classifications of the UV fixed point solutions of the couplings with relevant (Rev) and irrelevant (Irev) characteristics are listed. }
\label{UV fixed point_3}
\end{table}

\subsection{RG Flow}

To determine the RG flow of the system we can either consider a flow from the IR to the UV or vice-versa.  We shall primarily  focus on the UV to IR approach,  only briefly commenting on the IR to UV approach. 

For the UV to IR approach one simply starts from the UV fixed point and makes the RG flows run toward the IR. Here we use the fact that at one loop order, the beta functions of the three gauge couplings are completely decoupled from the other couplings. In addition, since the gauge couplings are UV relevant, we could freely choose the (IR) initial conditions as the gauge coupling matching conditions at a certain Trinification symmetry breaking scale.   We could therefore first solve the gauge coupling RG trajectories separately and then run the remaining couplings only along the determined RG trajectories of the gauge couplings. In principle, we could also run the RG flows along the alternative relevant coupling (e.g.~$\lambda_d$ in the symmetric case of Tab.~\ref{UV fixed point_1}). However, since $\lambda_d$ is not decoupled from the other couplings this would be more difficult to handle. 

\begin{figure}[t!]
\centering
\includegraphics[width=0.6\columnwidth]{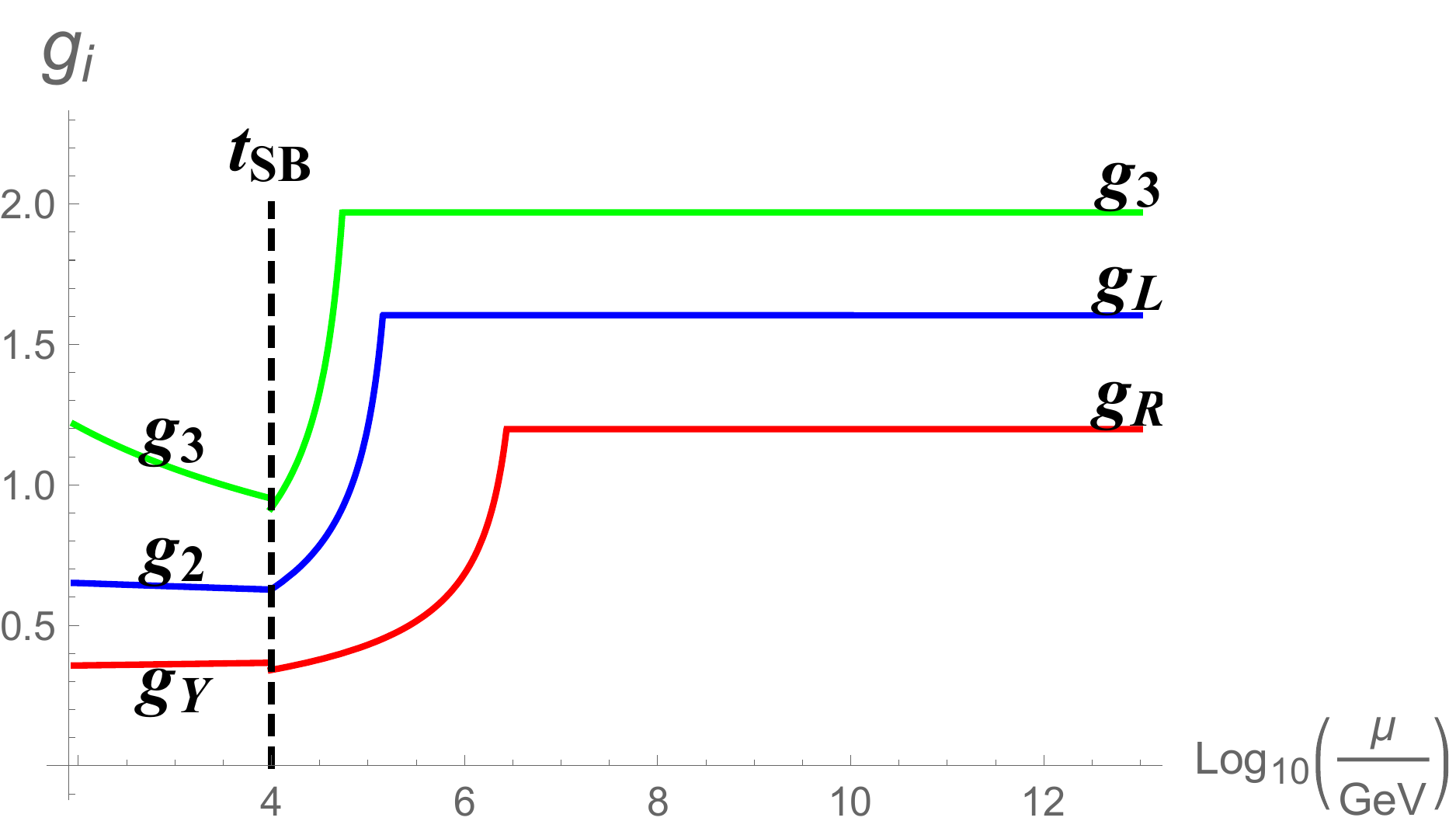}\hspace{0.06\columnwidth}
\caption{\small RG running of the gauge couplings by using the UV to IR approach. We have chosen $N_{FC}=95,\,N_{FL}=165,\,N_{FR}=62$. We  set the initial conditions for $g_L,\,g_R,\,g_c$ at the IR
used the matching conditions there (see Eq.~\eqref{matching_1}).
 For simplification, we have assumed that the vector-like fermions under gauge different symmetry groups are exactly introduced at the Trinification breaking scale, $t_{SB}=10\,\rm{TeV}$, marked by a vertical dashed line.}\label{Running_Couplings_1}
\end{figure}
In the IR to UV approach, the RG flow of the irrelevant couplings is constrained on the separatrices which are defined to divide the RG flow into distinct physical regions. We can therefore solve the set of equations $\beta_i=0$ ($i$ corresponding to all the irrelevant couplings) for all the irrelevant couplings as function of the relevant couplings. 
 We are thus free to choose the IR initial conditions of these relevant couplings to be compatible with the known phenomenological constraints while preserving safety at the UV scale.
The disadvantage of this approach  for a complicated system like   Trinification is that it is very hard to disentangle relevant and irrelevant couplings and analytically solve all the irrelevant couplings as function of the relevant couplings.

\begin{figure}[t!]
\centering
\subfigure[$y_{\psi_{Q1}}-\mu$]{
\begin{minipage}{7cm}
\centering
\includegraphics[width=0.8\columnwidth]{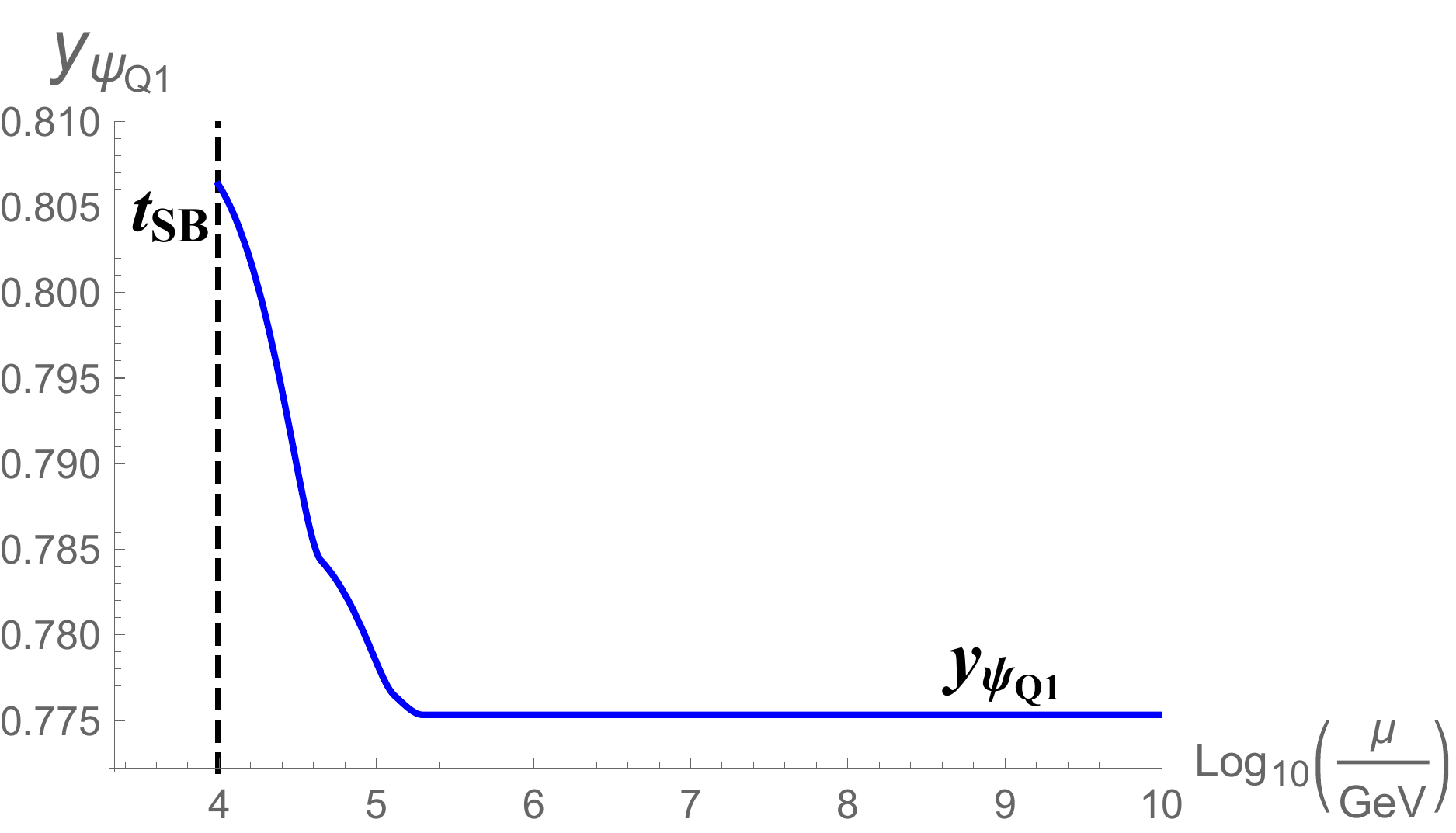}\hspace{0.06\columnwidth}
\end{minipage}
}
\subfigure[$y_{\psi_{E1}}-\mu$]{
\begin{minipage}{7cm}
\centering
\includegraphics[width=0.8\columnwidth]{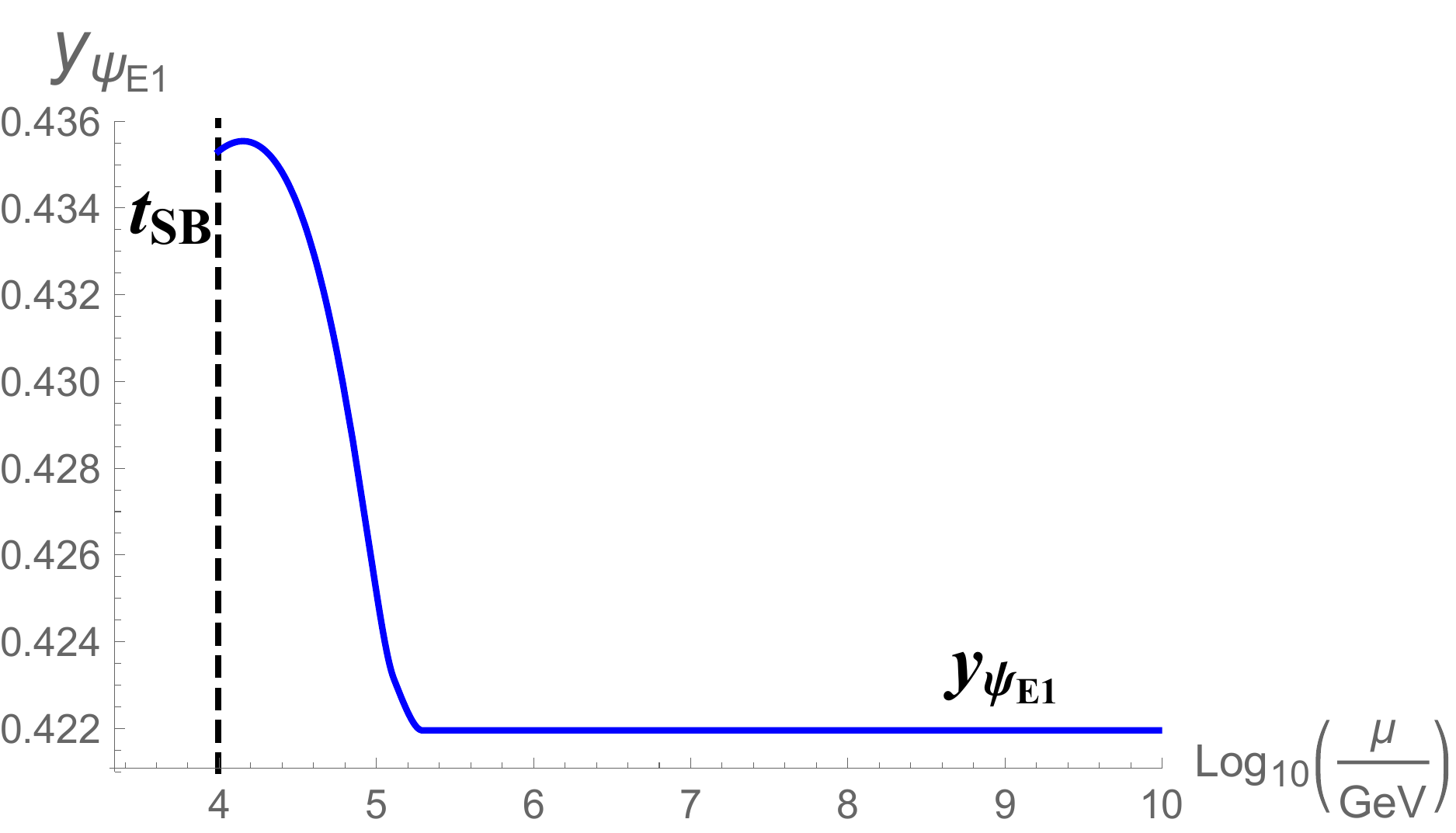}\hspace{0.06\columnwidth}
\end{minipage}
}
\subfigure[$y_{\psi_{Q2}}-\mu$]{
\begin{minipage}{7cm}
\centering
\includegraphics[width=0.8\columnwidth]{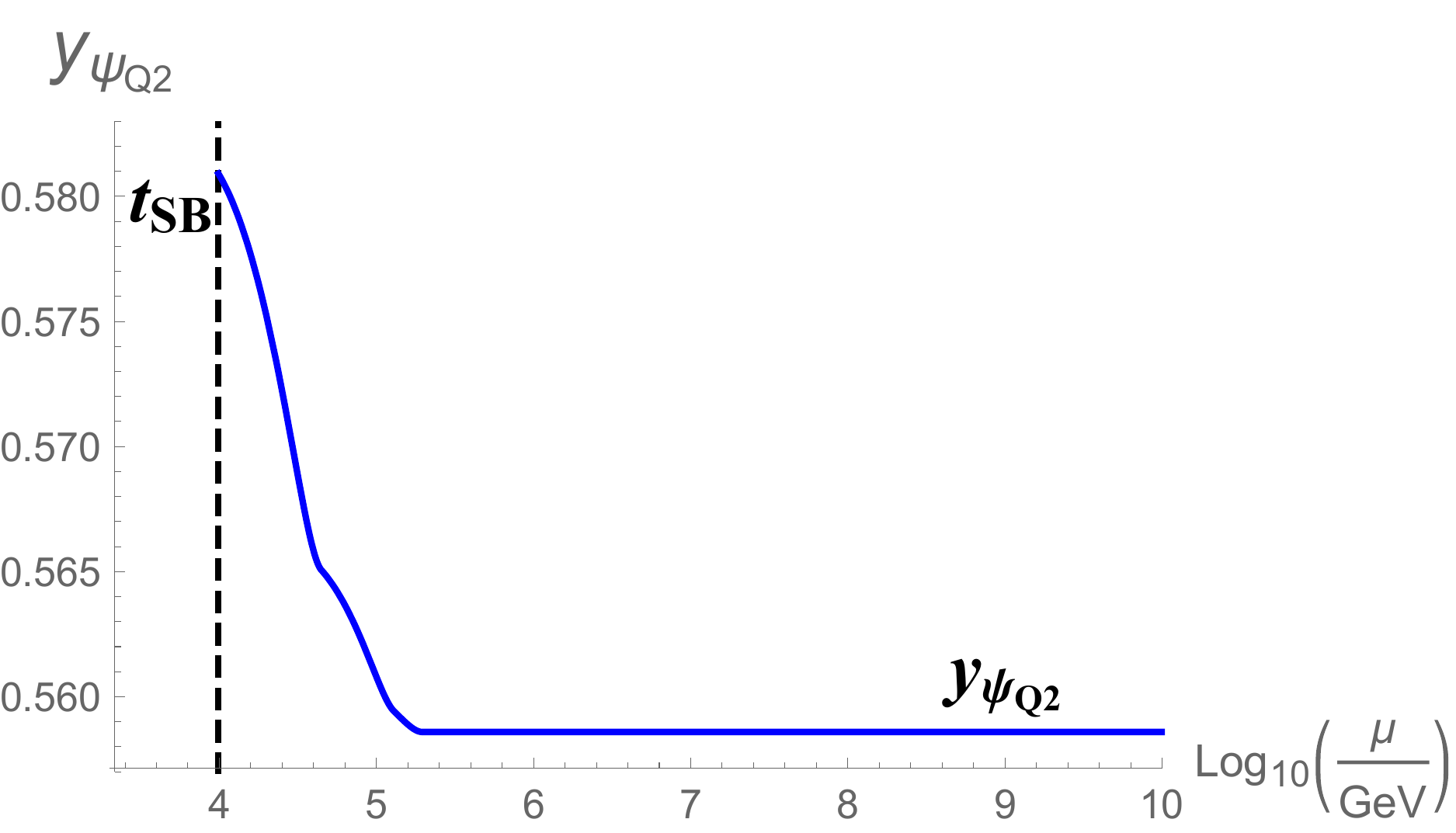}\hspace{0.06\columnwidth}
\end{minipage}
}
\subfigure[$y_{\psi_{E2}}-\mu$]{
\begin{minipage}{7cm}
\centering
\includegraphics[width=0.8\columnwidth]{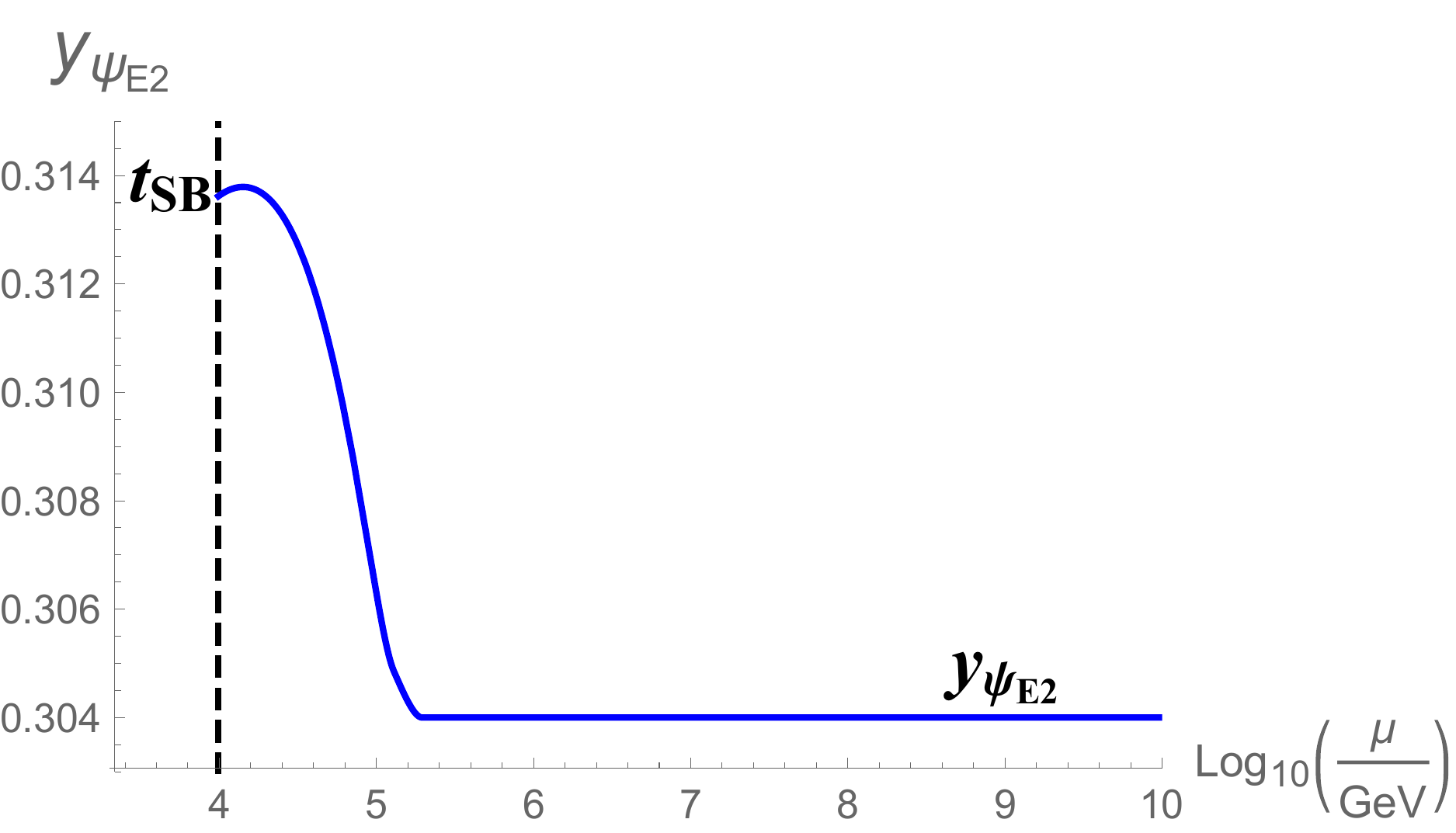}\hspace{0.06\columnwidth}
\end{minipage}
}
\caption{\small  RG running of the Yukawa couplings by using the UV to IR approach. We have chosen $N_{FC}=95,\,N_{FL}=165,\,N_{FR}=62$. For simplification, we have assumed that the vector-like fermions under gauge different symmetry groups are exactly introduced at the Trinification breaking scale, $t_{SB}=10\,\rm{TeV}$, marked by a vertical dashed line.}
\label{Running_Couplings_2}
\end{figure} 

\begin{figure}[htb]
\centering
\subfigure[$\lambda_{1a}-\mu$]{
\begin{minipage}{7cm}
\centering
\includegraphics[width=0.8\columnwidth]{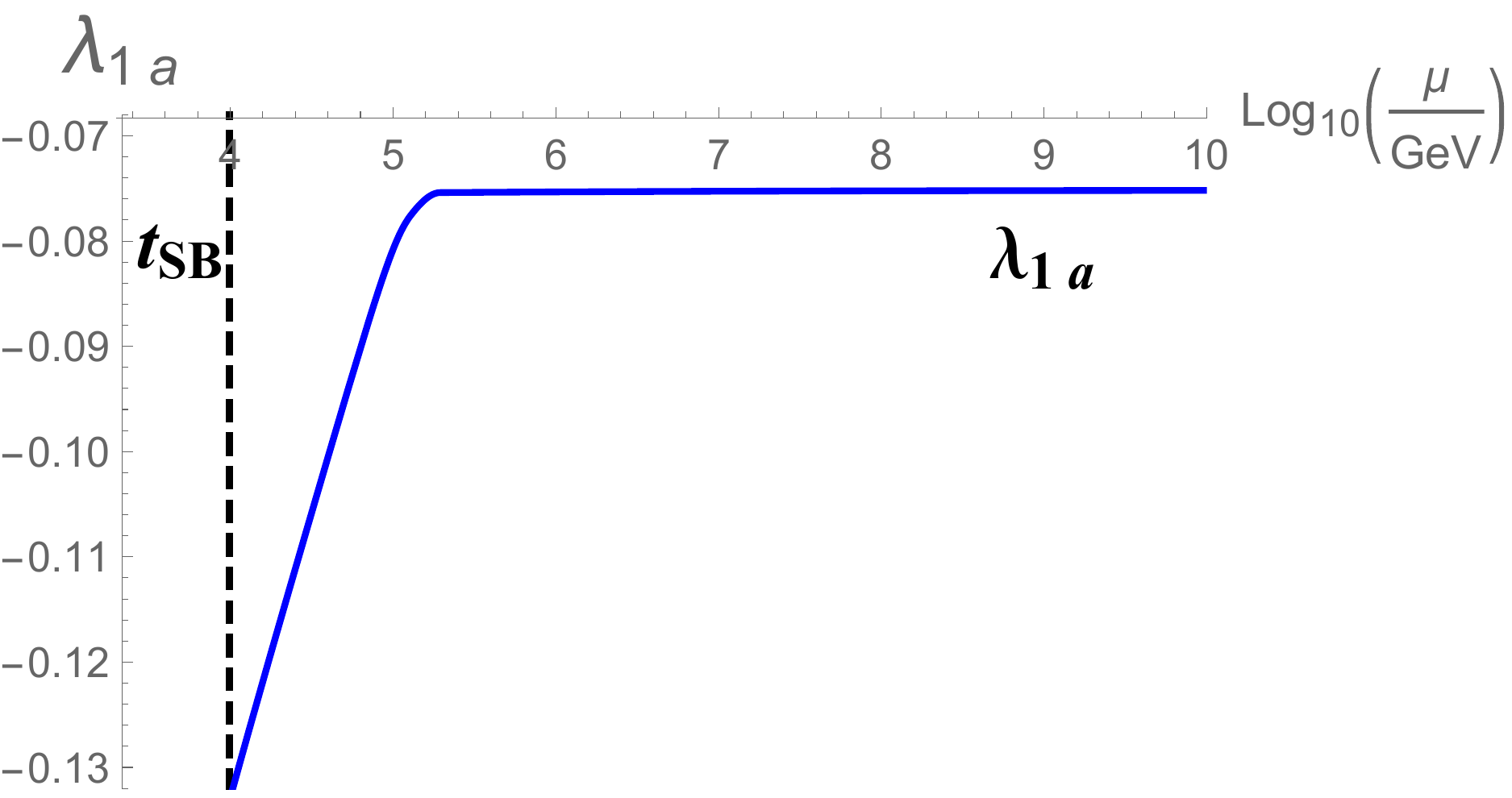}\hspace{0.05\columnwidth}
\end{minipage}
}
\subfigure[$\lambda_{1b}-\mu$]{
\begin{minipage}{7cm}
\centering
\includegraphics[width=0.8\columnwidth]{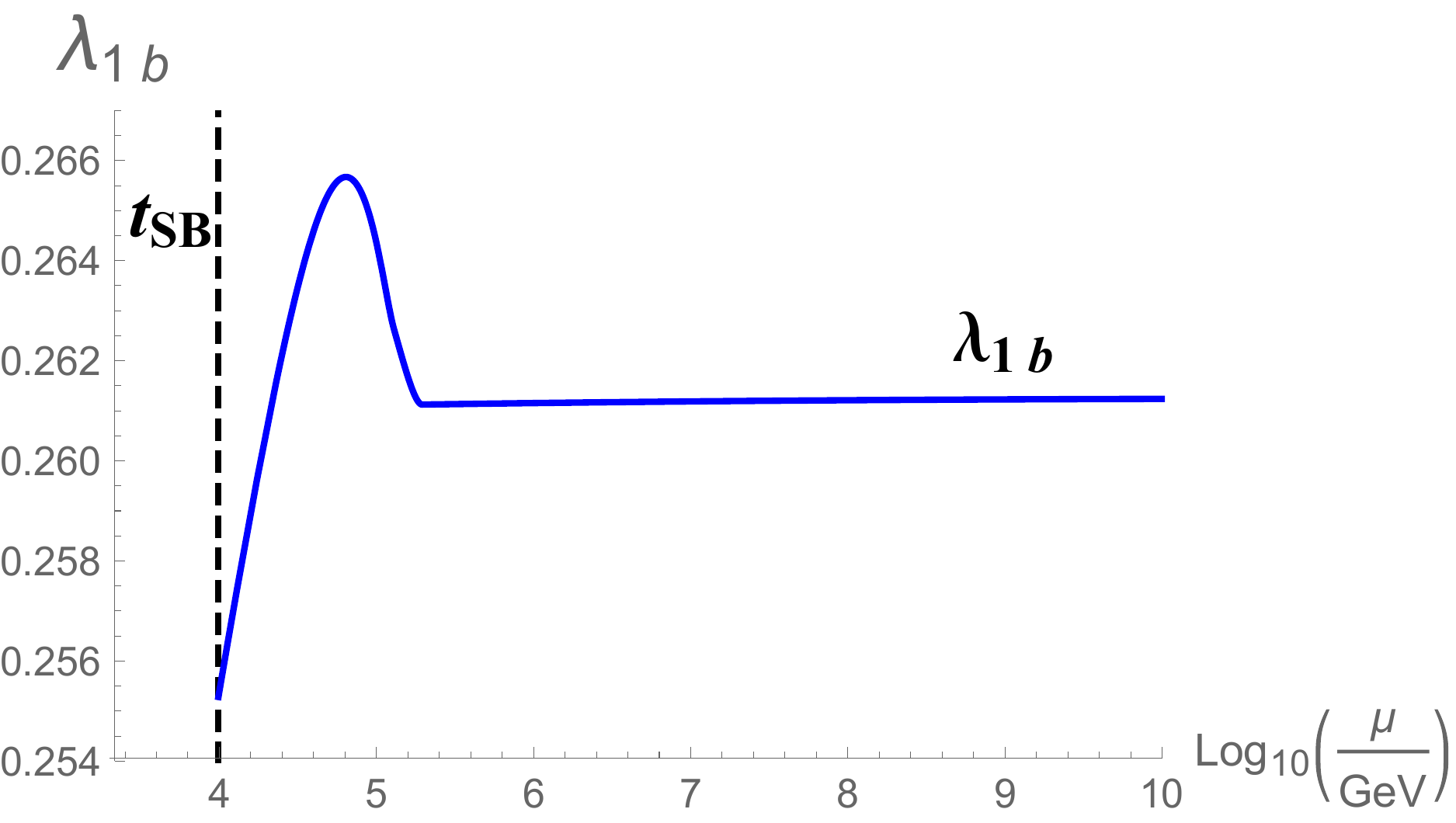}\hspace{0.05\columnwidth}
\end{minipage}
}
\subfigure[$\lambda_{2a}-\mu$]{
\begin{minipage}{7cm}
\centering
\includegraphics[width=0.8\columnwidth]{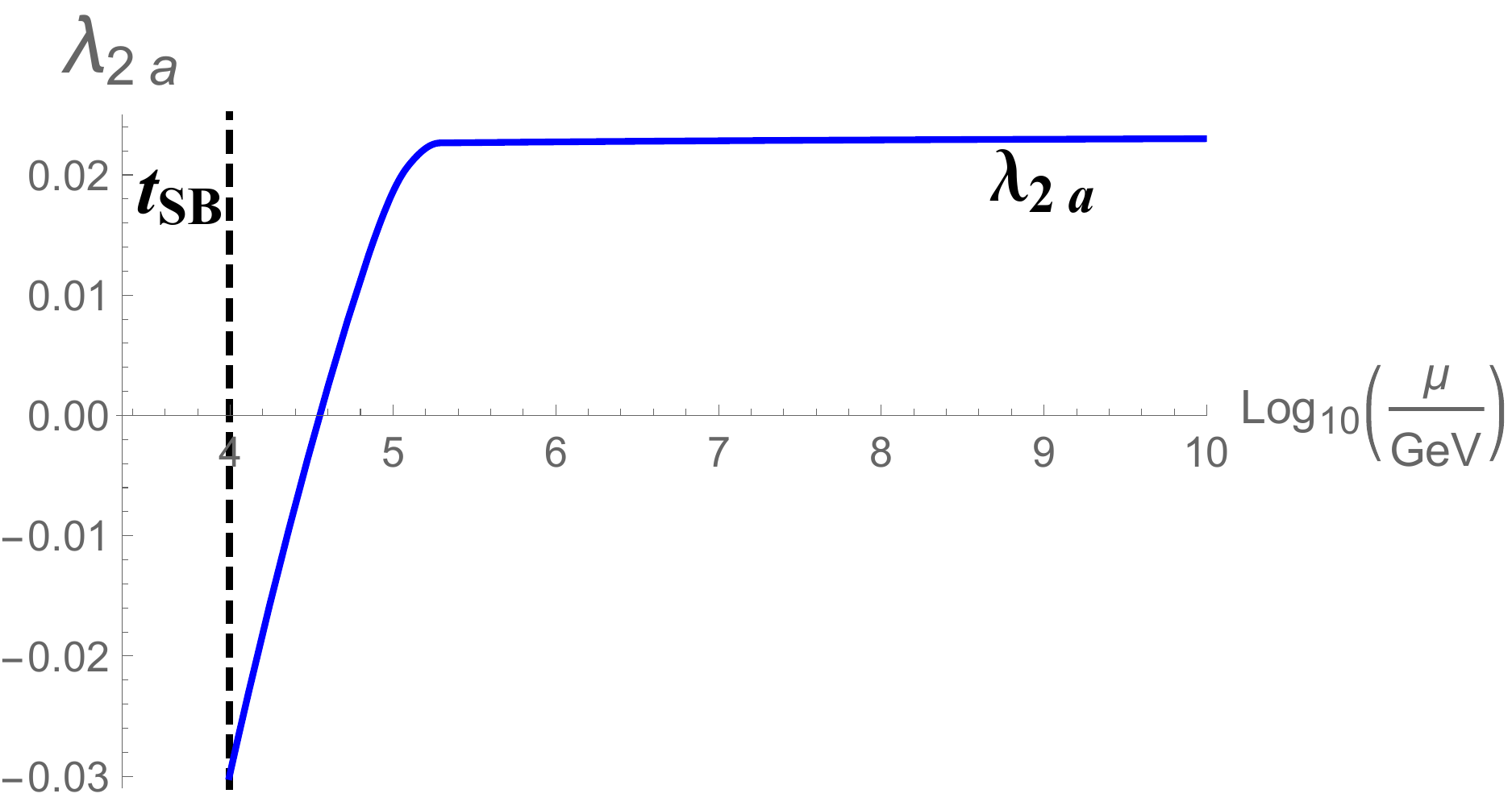}\hspace{0.05\columnwidth}
\end{minipage}
}
\subfigure[$\lambda_{2b}-\mu$]{
\begin{minipage}{7cm}
\centering
\includegraphics[width=0.8\columnwidth]{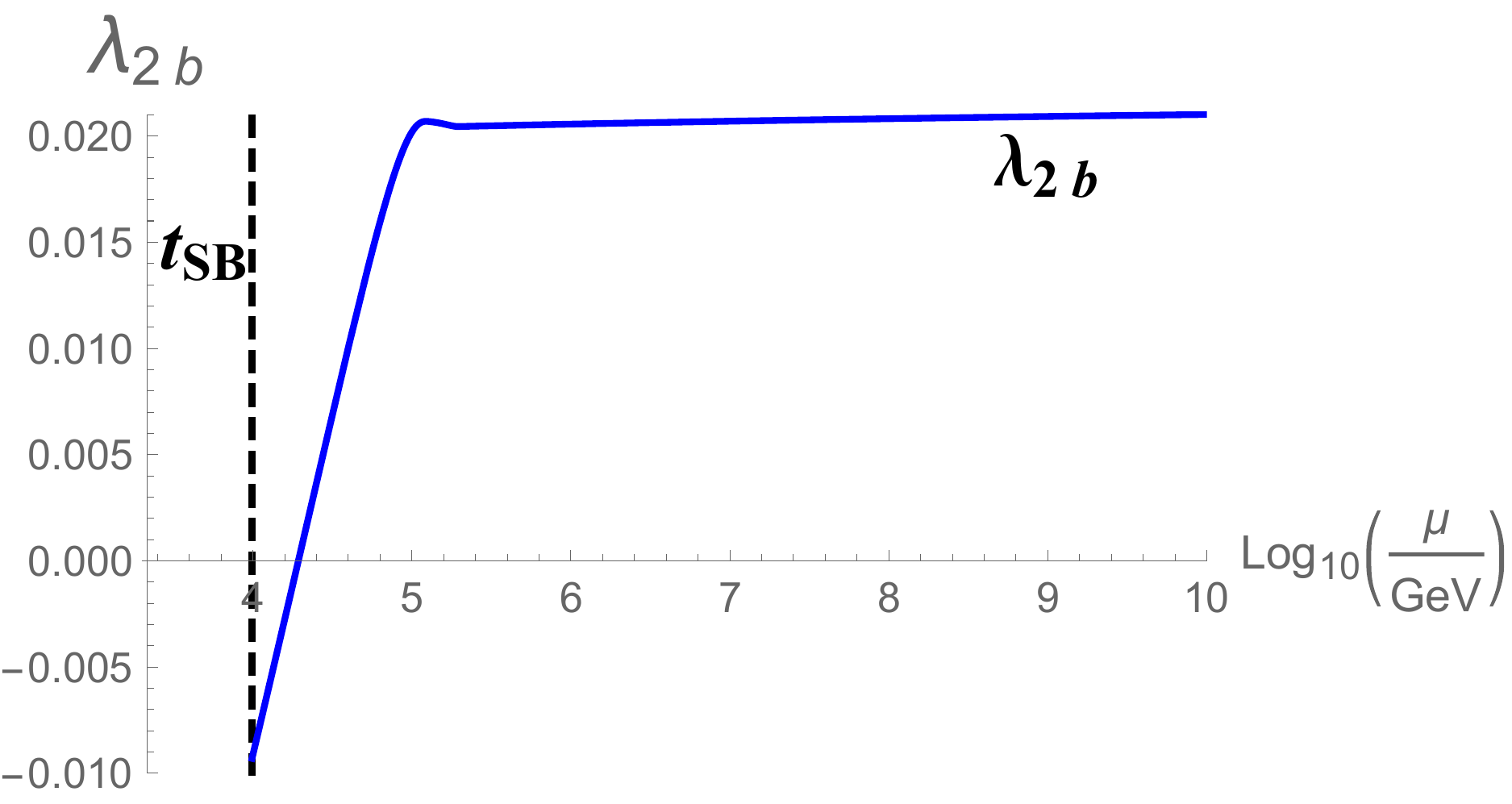}\hspace{0.05\columnwidth}
\end{minipage}
}
\caption{\small RG running of the scalar quartic couplings using the UV to IR approach for $N_{FC}=95,\,N_{FL}=165,\,N_{FR}=62$. All the vector-like fermions  appear (dashed line) at the symmetry breaking scale of the Trinification group, which is around  $10\,\rm{TeV}$.}
\label{Running_Couplings_3}
\end{figure} 

\begin{figure}[htb]
\centering
\subfigure[$\lambda_a-\mu$]{
\begin{minipage}{7cm}
\centering
\includegraphics[width=0.8\columnwidth]{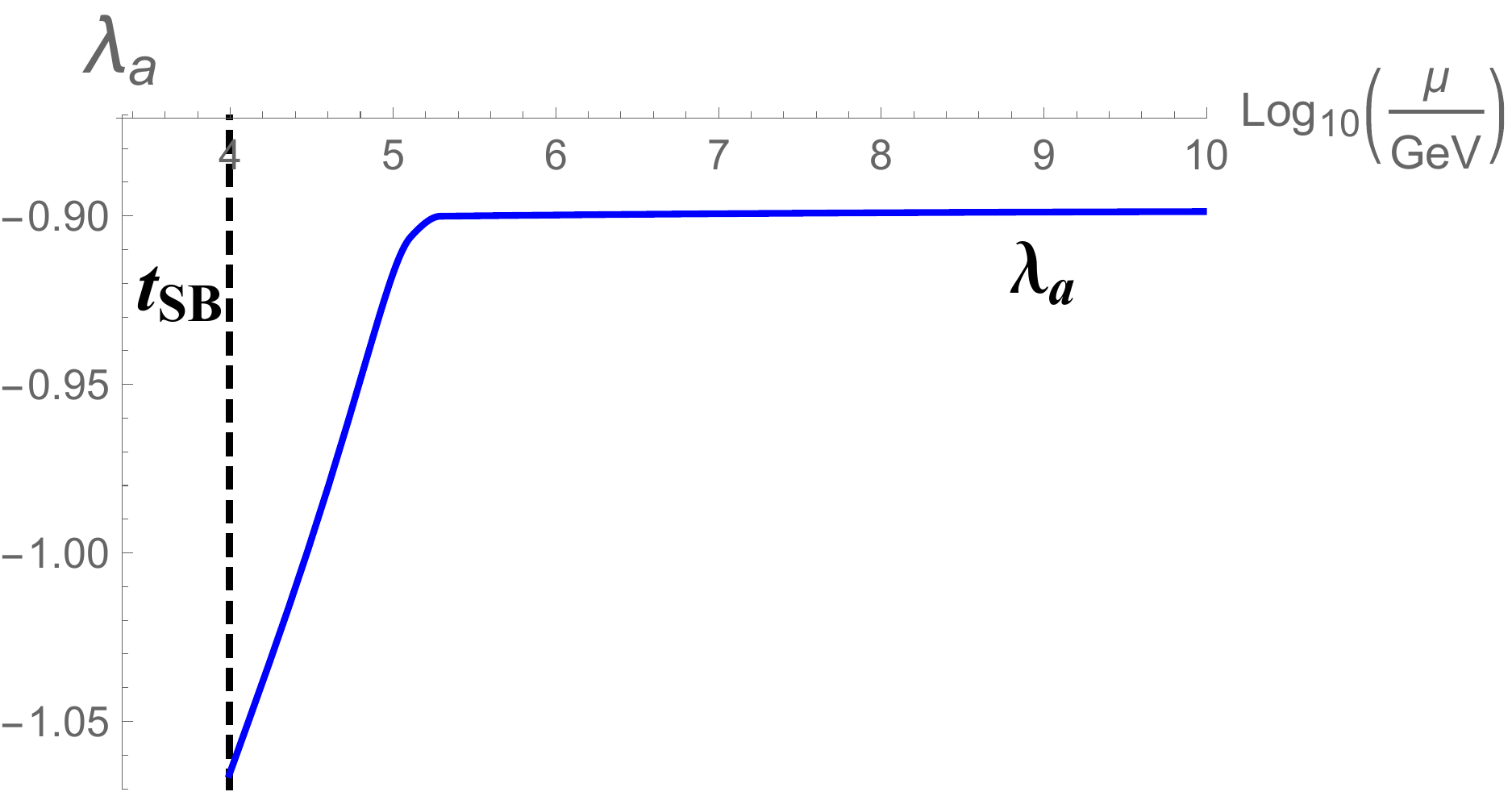}\hspace{0.05\columnwidth}
\end{minipage}
}
\subfigure[$\lambda_b-\mu$]{
\begin{minipage}{7cm}
\centering
\includegraphics[width=0.8\columnwidth]{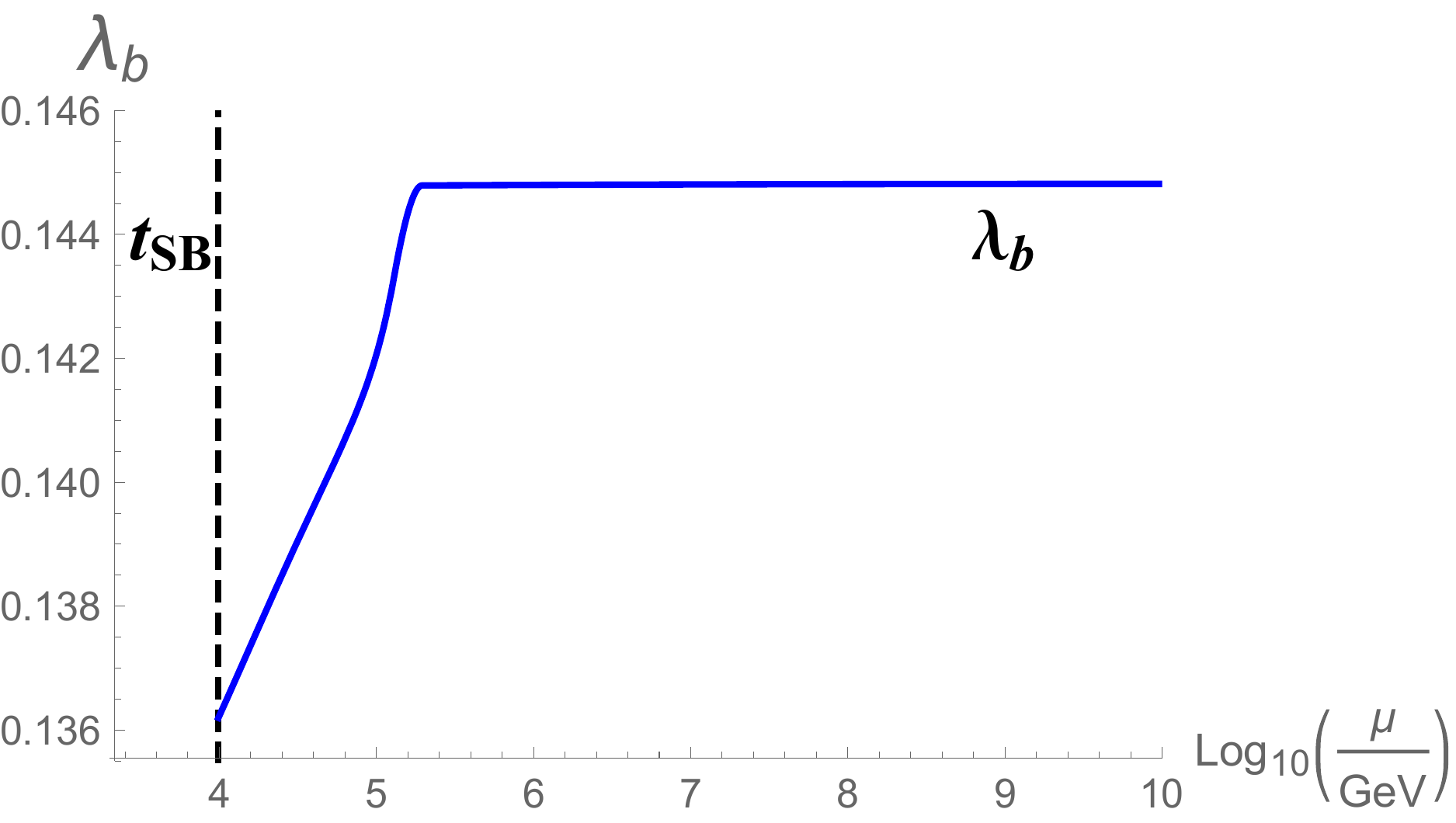}\hspace{0.05\columnwidth}
\end{minipage}
}
\subfigure[$\lambda_c-\mu$]{
\begin{minipage}{7cm}
\centering
\includegraphics[width=0.8\columnwidth]{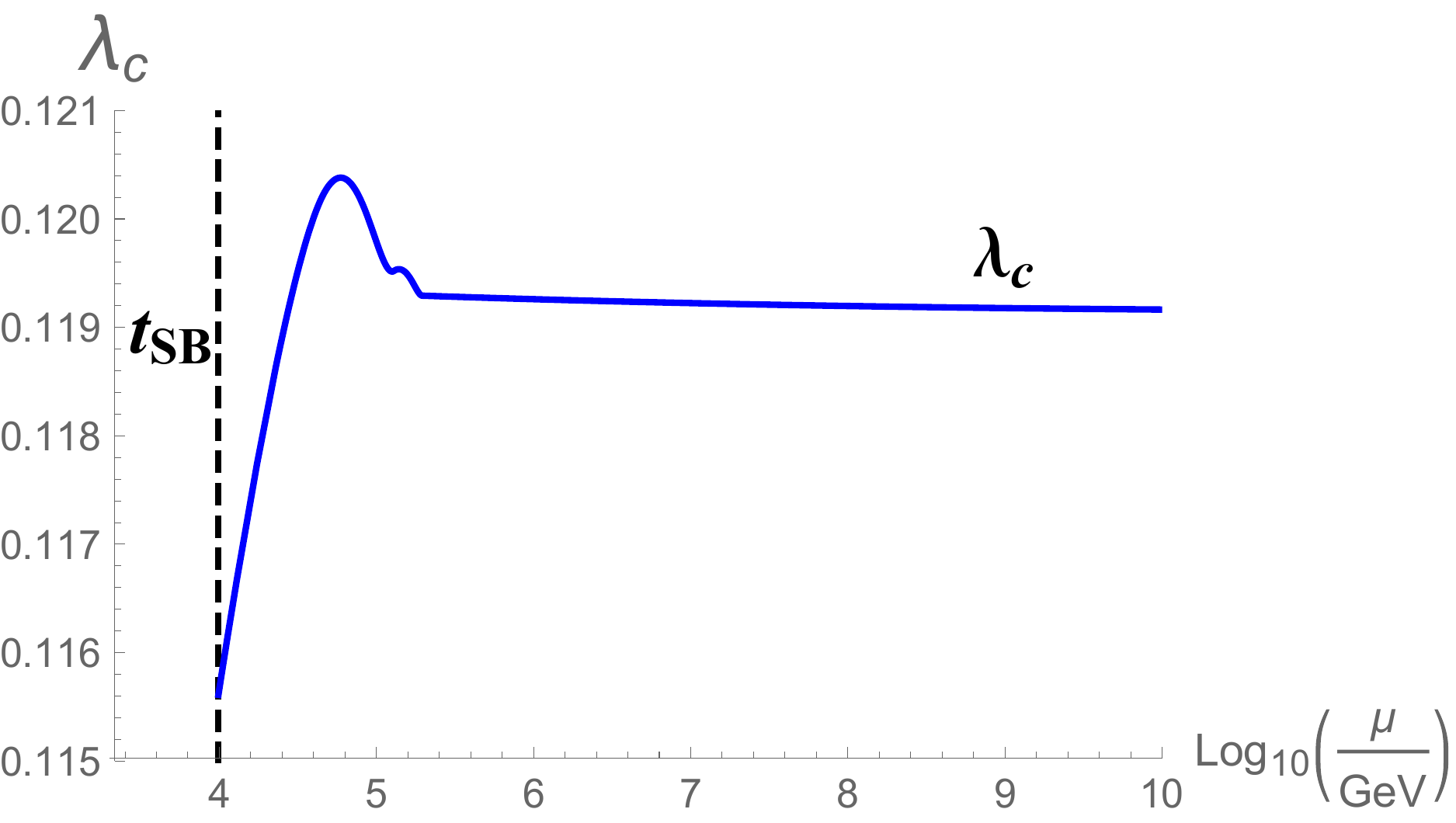}\hspace{0.05\columnwidth}
\end{minipage}
}
\subfigure[$\lambda_d-\mu$]{
\begin{minipage}{7cm}
\centering
\includegraphics[width=0.8\columnwidth]{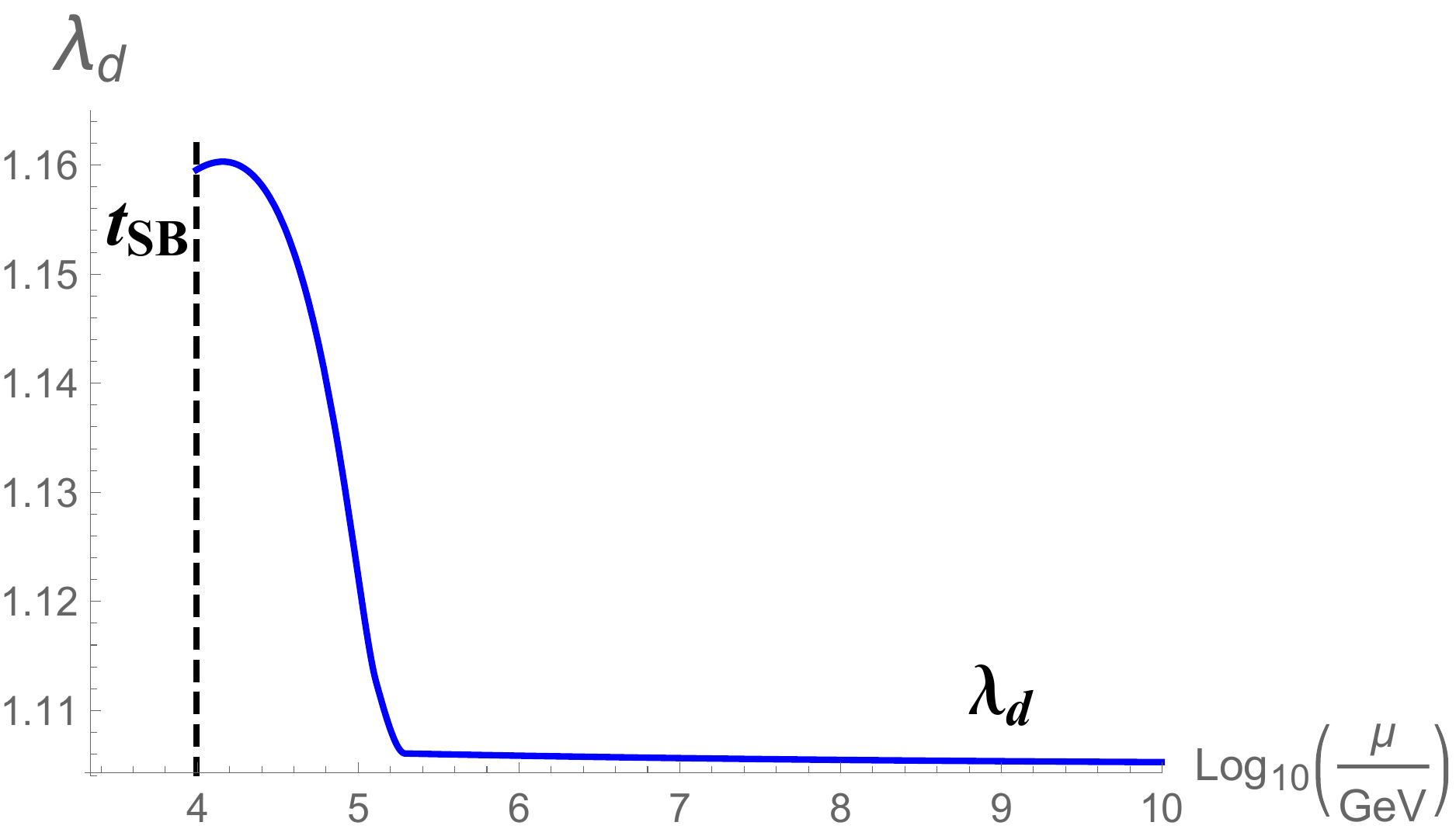}\hspace{0.05\columnwidth}
\end{minipage}
}
\subfigure[$\lambda_e-\mu$]{
\begin{minipage}{7cm}
\centering
\includegraphics[width=0.8\columnwidth]{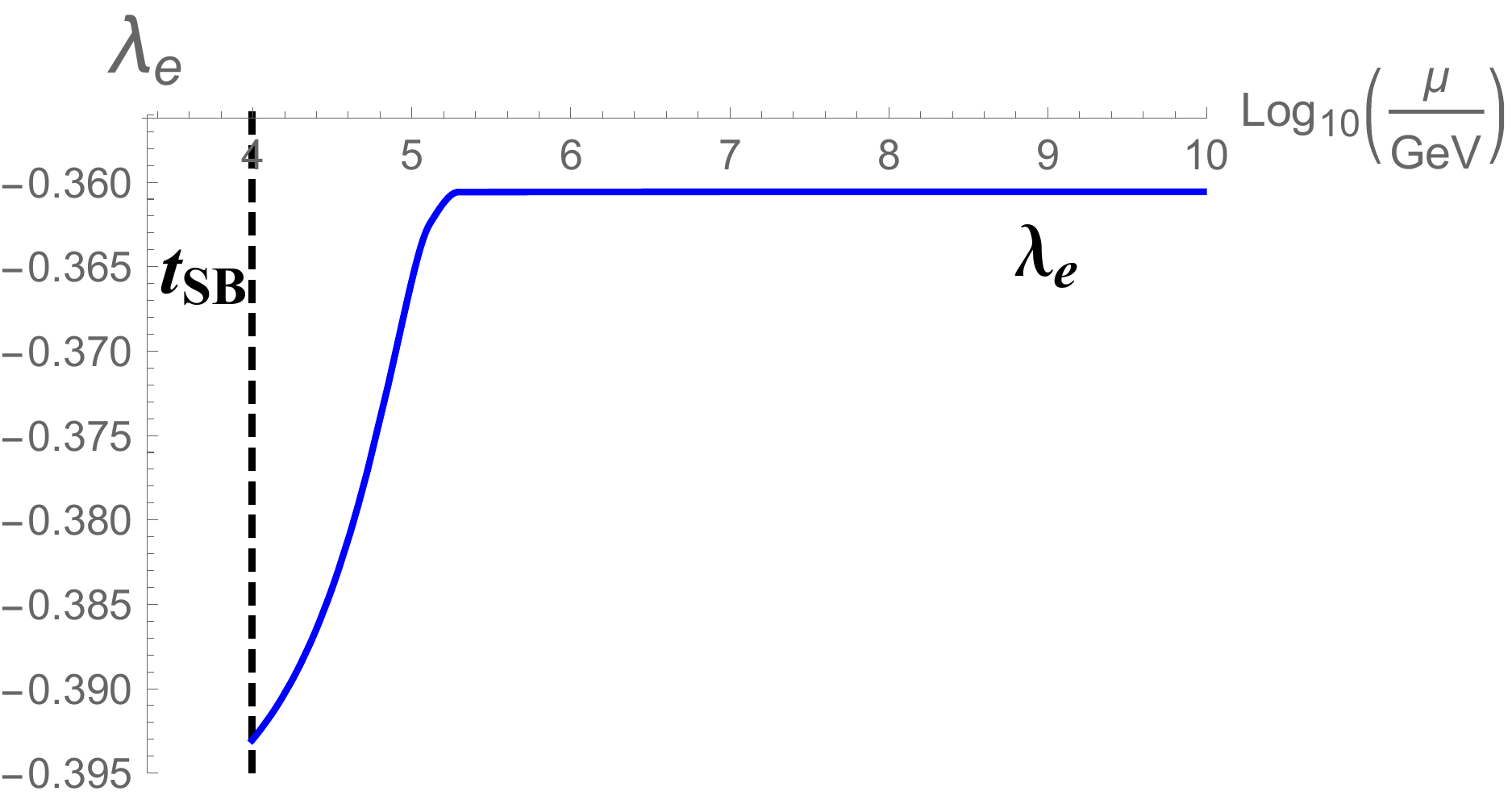}\hspace{0.05\columnwidth}
\end{minipage}
}
\subfigure[$\lambda_f-\mu$]{
\begin{minipage}{7cm}
\centering
\includegraphics[width=0.8\columnwidth]{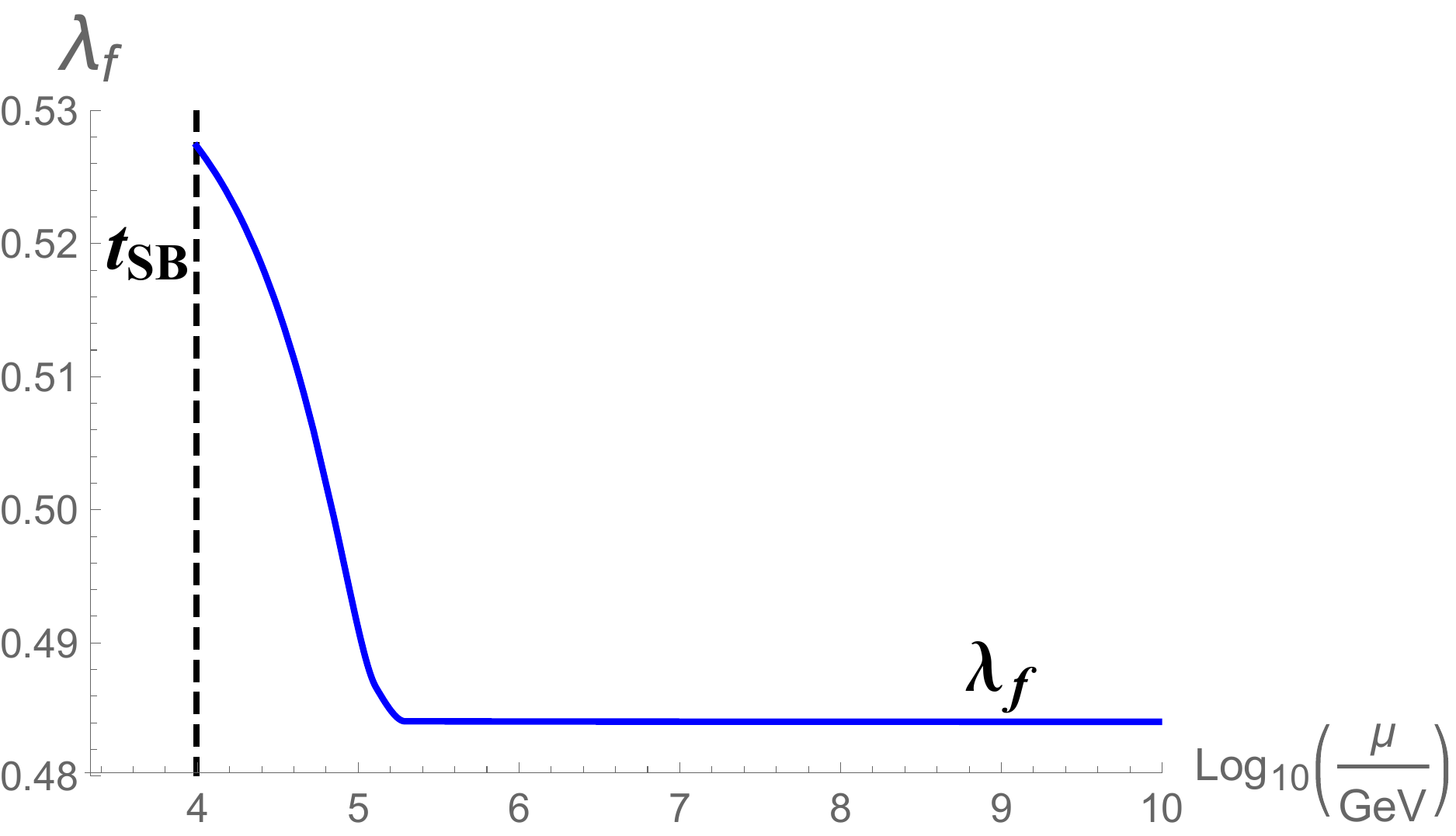}\hspace{0.05\columnwidth}
\end{minipage}
}
\caption{\small RG running of the scalar quartic couplings using the UV to IR approach for $N_{FC}=95,\,N_{FL}=165,\,N_{FR}=62$. All the vector-like fermions  appear at the symmetry breaking scale of the Trinification group, which is around  $10\,\rm{TeV}$ (dashed line).}
\label{Running_Couplings_4}
\end{figure} 

We report our results in Figs.~\ref{Running_Couplings_1}, \ref{Running_Couplings_2}, \ref{Running_Couplings_3}, \ref{Running_Couplings_4} where we show the running of the gauge, Yukawa and scalar couplings by using the UV to IR approach for  ($N_{FC}=95,\,N_{FL}=165,\,N_{FR}=62$). The corresponding UV fixed point solution is the one shown in   tables~\ref{UV fixed point_2} and~\ref{UV fixed point_3}. The RG flows of the gauge couplings are determined once the IR conditions are given as previously noted. The IR initial conditions for $g_L$, $g_R$ and $g_c$ are obtained by using the matching conditions~\eqref{matching_1} and the SM couplings  run from the EW scale to the Trinification symmetry breaking scale. For simplicity, the vector-like fermions masses are taken to be the Trinification symmetry breaking scale at $10\,\rm{TeV}$. From Figs.~\ref{Running_Couplings_1}, \ref{Running_Couplings_2}, \ref{Running_Couplings_3}, \ref{Running_Couplings_4}, it is clear that all couplings (i.e.~gauge, Yukawa and scalar quartic) achieve a safe UV fixed point. The transition scale, above which the UV fixed point is reached, is about $10^7\,\rm{GeV}$ for all the couplings.  Note that this transition scale is dependent on the number of vector-like fermions $N_F$. By increasing or decreasing $N_F$, the transition scales will correspondingly decrease or increase.

\section{Matching the Standard Model}\label{Sec4}

\subsection{Scalar Sector}

The low energy effective scalar field sector of the Trinification model contains four Higgs doublets. For simplification we shall assume the two Higgs doublets coming from the second scalar triplet play a less important role (decoupled)
at the electroweak scale (this corresponds to a special case where $n_1=n_2=0$ in Eq.~\eqref{scalar vacuum}) and  focus only on two of the four Higgs doublets. 

Thus, after Trinification symmetry breaking these scalar triplets should match the conventional two Higgs doublet model, which is defined as
\begin{equation}
\begin{split}
V_H&=m_{11}^2\Phi_1^\dagger\Phi_1+m_{22}^2\Phi_2^\dagger\Phi_2-\left(m_{12}^2\Phi_1^\dagger\Phi_2+\rm{H.c.}\right)\\
&+\frac{1}{2}\bar{\lambda}_1\left(\Phi_1^\dagger\Phi_1\right)^2+\frac{1}{2}\bar{\lambda}_2\left(\Phi_2^\dagger\Phi_2\right)^2+\bar{\lambda}_3\left(\Phi_1^\dagger\Phi_1\right)\left(\Phi_2^\dagger\Phi_2\right)+\bar{\lambda}_4\left(\Phi_1^\dagger\Phi_2\right)\left(\Phi_2^\dagger\Phi_1\right)\\
&+\left[\frac{1}{2}\bar{\lambda}_5\left(\Phi_1^\dagger\Phi_2\right)^2+\bar{\lambda}_6\left(\Phi_1^\dagger\Phi_1\right)\left(\Phi_1^\dagger\Phi_2\right)+\bar{\lambda}_7\left(\Phi_2^\dagger\Phi_2\right)\left(\Phi_1^\dagger\Phi_2\right)+\rm{H.c.}\right]\,.\label{two Higgs doublet}
\end{split}
\end{equation}
Comparing with \eqref{two Higgs doublet} with \eqref{triplet potential}, we find:
\begin{equation}
\begin{split}
&\bar{\lambda}_1=2\left(\lambda_{1a}+\lambda_{1b}\right),\quad\bar{\lambda}_2=2\left(\lambda_{1a}+\lambda_{1b}\right),\quad\bar{\lambda}_3=2\left(\lambda_{1a}+\lambda_{1b}\right),\quad\bar{\lambda}_4=-2\lambda_{1b}\\
&\bar{\lambda}_5=0,\quad\bar{\lambda}_6=0,\quad\bar{\lambda}_7=0\label{quartic_matching}\,,
\end{split}
\end{equation}
where the two Higgs doublet in the second triplet has been eliminated since it is much heavier and decoupled in our special scenario. The electroweak two Higgs doublet mainly comes from the first scalar triplet.

Given a set of values for $\left(N_{FC},\,N_{FL},\,N_{FR}\right)$  and a Trinification symmetry breaking scale,  we can obtain all the coupling values at the Trinification symmetry breaking scale in both the symmetric and asymmetric cases via RG running from the UV fixed point.
 Note that we will have multiple sets of possible values of the couplings at the symmetry breaking scale,  corresponding to different UV fixed point solutions since each set of choices for $\left(N_{FC},\,N_{FL},\,N_{FR}\right)$ provides multiple UV fixed point solutions. We could then use the coupling values obtained at the Trinification symmetry breaking scale as  new initial conditions (also implementing the matching conditions~\eqref{quartic_matching}). Eventually, by using the two Higgs doublet RG beta functions \cite{Branco:2011iw}, we could obtain all the coupling values at the electroweak scale.

In addition, the mass matrix (neutral scalar fields) of the two Higgs doublet model (setting $m_{12}^2\rightarrow0$ for simplificity) is given by:
\begin{equation}
M^2_{\rm{neutral}}=\left[
\begin{array}{cc}
 2 \bar{\lambda}_1 v_1^2 & \left(\bar{\lambda} _3+\bar{\lambda}_4+\bar{\lambda} _5\right) v_1 v_2 \\
 \left(\bar{\lambda}_3+\bar{\lambda} _4+\bar{\lambda}_5\right) v_1 v_2 & 2 \bar{\lambda} _2 v_2^2 \\
\end{array}
\right]
\end{equation}

Note that this mass matrix is scale dependent and is defined at the electroweak scale. By using the coupling values obtained previously at the electroweak scale, we could calculate the mass eigenvalues. We expect both eigenvalues of the mass matrix should be positive and the smaller of the two eigenvalues to be close to the $125\,\rm{GeV}$ Higgs mass. It can be shown that in the asymmetric case, by choosing $N_{FC}=95,\,N_{FL}=165,\,N_{FR}=62$, we obtain: 
\begin{equation}
\bar{\lambda}_1=0.234,~\bar{\lambda}_2=0.331,~\bar{\lambda}_3=0.213,~\bar{\lambda}_4=-0.499,~\bar{\lambda}_5=0,~y_{\psi_{Q1}}=0.806,~y_{\psi_{E1}}=0.435\,,\label{couplingvalue_EW}
\end{equation}
which leads to two neutral scalars with masses  $125.4\,\rm{GeV}$ for the lighter Higgs and $765.4\,\rm{GeV}$ for the heavier one.  The quartic couplings are at around the electroweak scale while the Yukawa couplings are at the Trinification symmetry breaking scale $\sim 10\,\rm{TeV}$.

\begin{figure}[t!]
\centering
\includegraphics[width=0.4\columnwidth]{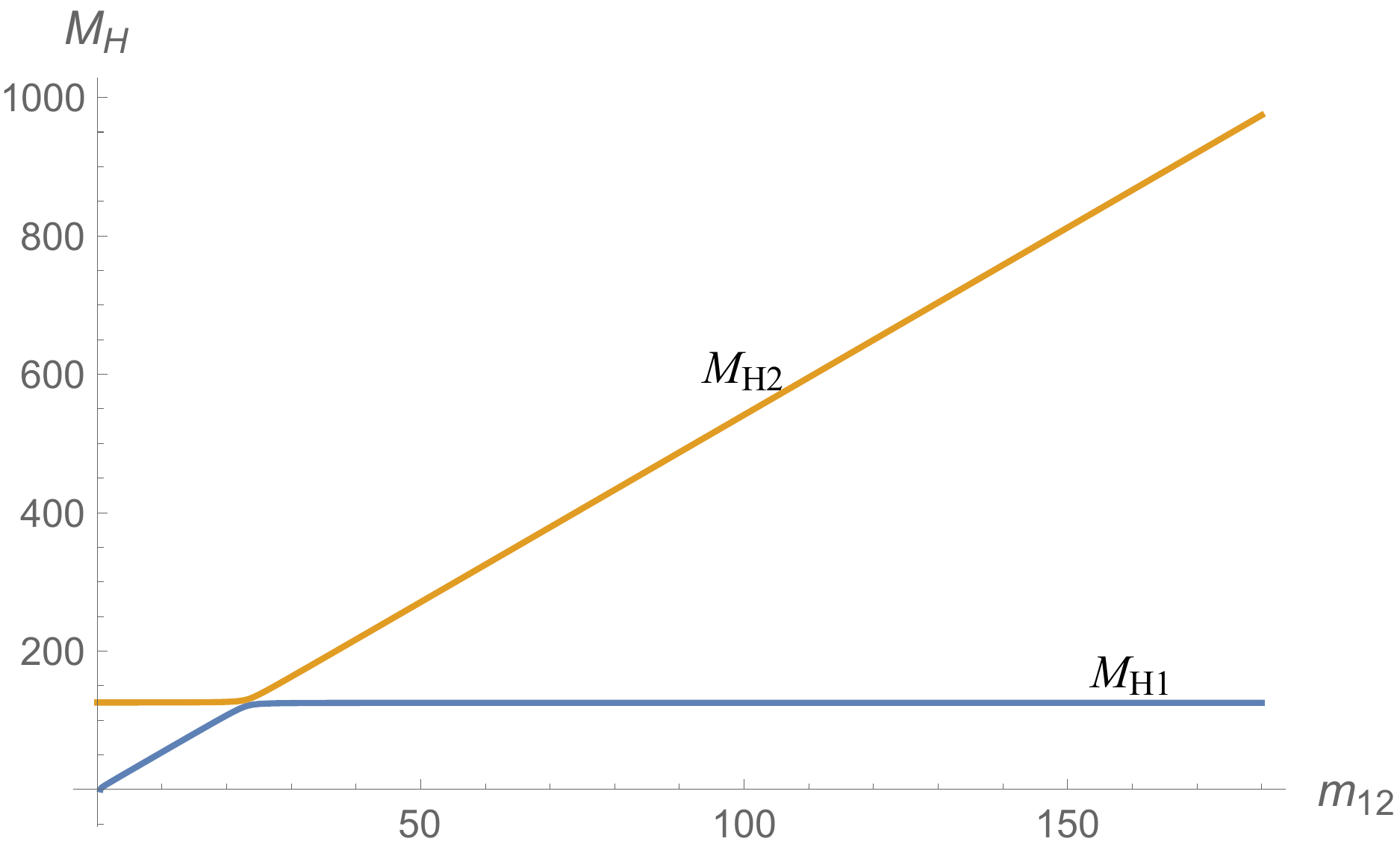}\hspace{0.04\columnwidth}
\caption{\small In this figure, we present two mass eigenvalues of the CP-even neutral Higgs mass matrix as a function of $m_{12}$.  $M_{H1}$ denotes the lighter Higgs while $M_{H2}$ denotes the heavier one.}
\label{mass eigenvalues}
\end{figure}

It is interesting to note that the above conclusion is dependent on the choice of the $m_{12}$ mass parameter in~\eqref{two Higgs doublet}. When setting $m_{12}=0$, the light Higgs will be massless while the heavy Higgs will be around $125\,\rm{GeV}$. However for a slight increase in the $m_{12}$ parameter, the light Higgs will increase correspondingly (see fig.~\ref{mass eigenvalues}) until $m_{12}\sim25\,\rm{GeV}$. After that the light Higgs mass is frozen at around $125\,\rm{GeV}$, whereas the heavy Higgs mass keep increasing with increasing $m_{12}$. More interestingly, the Trinification symmetry breaking scale cannot be too much smaller than $10\,\rm{TeV}$ or the prediction of the light Higgs mass will be a little bit too small, providing a dynamical explanation of both the symmetry breaking scale and why  new physics has not yet been found. 

Comparison of the Trinification results with the Pati-Salam case is also warranted. In the Pati-Salam case, in order to match the correct Higgs mass at the electroweak scale a constraint is placed on the Pati-Salam symmetry breaking scale that it not be  below $10000\,\rm{TeV}$ \cite{Molinaro:2018kjz}. This is much higher than the Trinification scale obtained here and so asymptotically safe Trinification theory is considerably more amenable to experimental tests.

The reason that a lower viable symmetry breaking in the Trinification model is obtain could be as follows. A larger Higgs mass prediction requires a large scalar quartic coupling at the electroweak scale. However, there are only two ways that could yield a large scalar quartic coupling at the electroweak scale. One is that of increasing the symmetry breaking scale, which will provide a longer  scale running distance for the quartic coupling to be increased by the top Yukawa coupling.  The other is to  increase the  value of the quartic coupling at the fixed point. This quantity is further determined by the quartic beta function, which is balanced by the contributions of the gauge couplings. Hence the larger the gauge coupling contributions at the quartic beta functions, the larger the Higgs mass at the electroweak scale. Fortunately the gauge couplings in the Trinification model have larger contributions relative to the Pati-Salam model simply due to the group structure. 

It can be shown that in the symmetric case, we are able to obtain  coupling solutions with features similar  to the asymmetric case. However to obtain a reasonable (sufficiently large) Higgs mass, the Trinification symmetry breaking scale in the symmetric case is required to be much larger (say around $100\,\rm{TeV}$), making the symmetric case somewhat less experimentally accessible.
By choosing $N_{FC}=92,\,N_{FL}=182,\,N_{FR}=38$, we obtain
\begin{equation}
\bar{\lambda}_1=0.067,~~\bar{\lambda}_2=0.336,~~\bar{\lambda}_3=0.056,~~\bar{\lambda}_4=0.089,~~\bar{\lambda}_5=0\,,\label{couplingvalue2_EW}
\end{equation}
which yields  two neutral scalars with masses  $126.6\,\rm{GeV}$ for the lighter Higgs and $541.3\,\rm{GeV}$ for the heavier one,  choosing $m_{12}\sim100\,\rm{GeV}$.

\subsection{Yukawa Sector}
The low energy effective field theory of the Trinification model is the type II two Higgs doublet model in which one scalar field couples only to the up-type quarks and the other couples to the down-type quarks and leptons. The Yukawa sector of the two Higgs doublet model can be written as \cite{Grinstein:2013npa}:
\begin{equation}
\label{yukawacouplings}
	\mathcal{L}_{yuk} = -\frac{M_u}{v}\left(\frac{\cos\tilde{\alpha}}{\sin\tilde{\beta}}\right) \bar u u h - \frac{M_d}{v}\left(\frac{\sin\tilde{\alpha}}{\cos\tilde{\beta}}\right) \bar d d h - \frac{M_e}{v}\left(\frac{\sin\tilde{\alpha}}{\cos\tilde{\beta}}\right) \bar e e h\,,
\end{equation}
where $\tilde{\alpha},\,\tilde{\beta}$ denote respectively the mixing angles of the two neutral CP-even  and CP-odd Higgs states, with $\tan\tilde{\beta}=\frac{u_2}{u_1}$ and $\vert\tilde{\alpha}-\tilde{\beta}\vert\sim\pi/2$. For simplicity, in Eq.~\eqref{yukawacouplings}, we have only written down explicitly the terms with only the $125\,\rm{GeV}$ light scalar state $h$. Comparing with Eqs.~\eqref{top mass} and \eqref{neutrino_mass_tree} with Eq.~\eqref{yukawacouplings}, we obtain the following relationships between the Standard Model Yukawa couplings and the Yukawa couplings of the Trinification model:
\begin{equation}
y^{\rm{SM}}_{\rm{top}}=y_{\psi_{Q1}}\cos\tilde{\alpha},\qquad y^{\rm{SM}}_{\rm{bottom}}=y_{\psi_{Q1}}\sin\alpha\sin\tilde{\alpha},\qquad y^{\rm{SM}}_{\rm{tau}}=y_{\psi_{E1}}\sin\beta\sin\tilde{\alpha}\,,\label{Yukawa matching}
\end{equation}
where $y^{\rm{SM}}_{\rm{top}},\,y^{\rm{SM}}_{\rm{bottom}},\,y^{\rm{SM}}_{\rm{tau}}$ denote respectively the SM Yukawa couplings of the top quark, bottom quark and tau lepton. We find in the asymmetric case, by choosing $N_{FC}=95,\,N_{FL}=165,\,N_{FR}=62$, we obtain 
\begin{equation}
\sin\alpha\sim\sin\beta\sim\sqrt2/2,\qquad u_1=8.4\,\rm{GeV},\qquad u_2=245.86\,\rm{GeV}\label{angle predicted}
\end{equation}
where we have used Eq.~\eqref{angle relationship} to provide a reasonable neutrino mass. Using Eqs.~\eqref{couplingvalue_EW}\eqref{Yukawa matching}\eqref{angle predicted},
we obtain:
\begin{equation}
\begin{split}
y^{\rm{Pre}}_{\rm{top}}&=0.806,\qquad y^{\rm{Pre}}_{\rm{bottom}}=0.019,\qquad y^{\rm{Pre}}_{\rm{tau}}=0.011\\
y^{\rm{SM}}_{\rm{top}}&=0.780,\qquad y^{\rm{SM}}_{\rm{bottom}}=0.019,\qquad y^{\rm{SM}}_{\rm{tau}}=0.008\,,
\end{split}
\end{equation}
where the first line denotes the Yukawa coupling predictions from the safe Trinification  model at the symmetry breaking scale $10\,\rm{TeV}$, while the second line denotes the SM Yukawa coupling at the same scale for comparison. It is clear that both top and bottom Yukawa couplings match extremely well. The tau lepton has a $27\%$ difference, which might be addressed by a more careful RG running procedure such as including the threshold contributions.

\subsection{Overview of the Parameter Space}

\begin{figure}[t!]
\centering
\includegraphics[width=0.6\columnwidth]{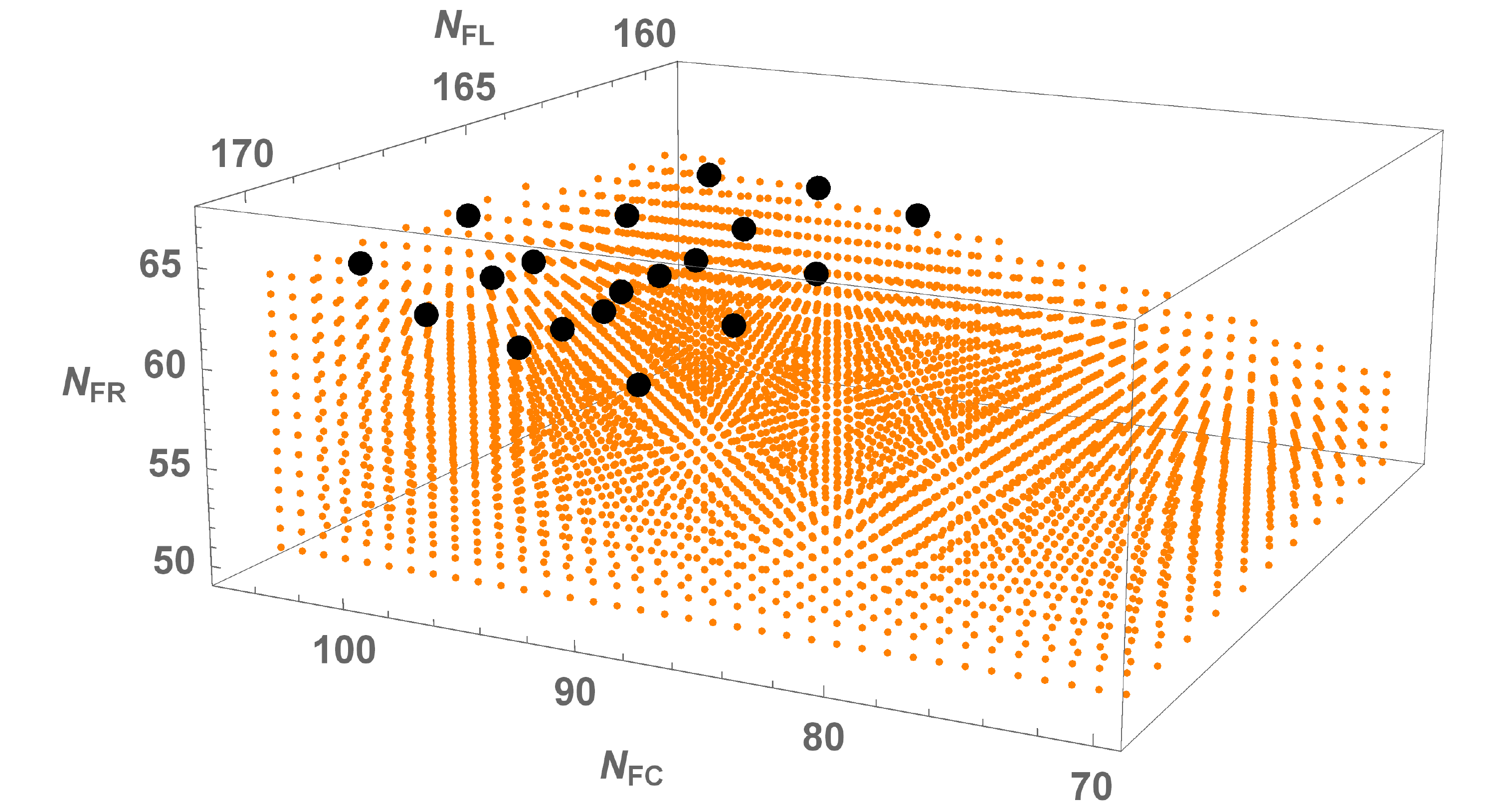}\hspace{0.04\columnwidth}
\caption{\small In this figure, we show the 3D parameter space constructed by the values of $N_{FC}\in\left(70,\,105\right),\,N_{FL}\in\left(160,\,170\right),\,N_{FR}\in\left(50,\,68\right)$ in the asymmetric case. The orange dot corresponds to a UV fixed point solution which does not match the SM at IR while the black dot corresponds to the viable UV fixed point solutions.}
\label{3D scan}
\end{figure}
In this section, we try to provide an intuitive picture of the viable parameter space. A scan of the whole parameter space, where $N_{FC},\,N_{FL},\,N_{FR}$ all range from $50$ to $200$, indicates that  the only viable region (for which the solutions could match the SM) is   where $N_{FC}\in\left(88,\,105\right),\,N_{FL}\in\left(160,\,170\right),\,N_{FR}\in\left(58,\,68\right)$. 

In fig.~\ref{3D scan}, we show the 3D parameter space constructed by the values of $N_{FC}\in\left(70,\,105\right),\,N_{FL}\in\left(160,\,170\right),\,N_{FR}\in\left(50,\,68\right)$ where
the orange dots correspond to  UV fixed point solutions that do not match the SM at the IR. The black dots correspond to  viable UV fixed point solutions. It is clear that of the many UV fixed point solutions only very few can roughly match the SM in the IR. We also note that when $N_{FC}$ gets smaller the white region, which corresponds to the parameter space without any fixed point solutions, gets much larger.
 
\section{Conclusions}
\label{Sec5}

A truly fundamental theory requires the presence of scale invariance at short distances \cite{Wilson:1971bg,Wilson:1971dh,Antipin:2013exa,Steele:2013fka,Wang:2015sxe,Wang:2015cda,Pelaggi:2017wzr,Shaposhnikov:2018jag}.
Fundamentality and naturality are complementary concepts. Short distance scale invariance implies fundamentality while (near) long distance conformality and/or  controllably broken symmetries (e.g.~Coleman Weinberg mechanism) help with naturality \cite{Antipin:2013exa,Steele:2013fka,Wang:2015sxe,Wang:2015cda,Pelaggi:2017wzr}.

We have here constructed a realistic asymptotically safe Trinification model in which the SM is embedded. Employing  large $N_F$ techniques, we demonstrated that all the couplings (i.e.~the gauge, scalar quartic and Yukawa couplings) achieve a UV interacting  fixed point far below the Planck scale. Different from the conventional GUT scenario, the unification of {\it all} type of couplings occurs due not only to a symmetry principle but also to 
 a dynamical principle, namely the presence of a UV fixed point.  We emphasize that we have shown that starting from a UV fixed point, a few RG flows can nicely match both the SM Higgs mass, the Yukawa couplings of the top and bottom quarks, and the reasonable neutrino masses, implying a truly UV completion of the Standard Model. The very few viable solutions in a 3D scan  of the whole parameter space is indicative of the highly predictive power of   asymptotically safe theories, which significantly narrow the parameter space of the theory at IR. 

It is also very intriguing that the Trinification symmetry breaking scale cannot be too much smaller than $10\,\rm{TeV}$ in order to procure a viable $125\,\rm{GeV}$ light Higgs mass. In comparison with  the $10000\,\rm{TeV}$ symmetry breaking scale in the asymptotically safe Pati-Salam model, the Trinification model is much more vulnerable to experiment.

\section*{Acknowledgments}
The work is supported by the Natural Sciences and Engineering Research Council of Canada (NSERC) and Danish National Research Foundation under the grant DNRF:90. Z.W.~Wang thanks Francesco Sannino, Tom Steele, Bob Holdom, Steven Abel, Wei-Chih Huang and Chen Zhang for very helpful suggestions.

\appendix 
\section{One-Loop RG equations of the Trinification model}\label{RG_all}
\label{App1}
\subsection{Gauge couplings}
\begin{equation}
\begin{split}
\left(4\pi \right)^2\frac{dg_{L}}{d\ln \mu}&=-\left(5+\frac{n_H}{2}\right)g_{L}^2\\
\left(4\pi \right)^2\frac{dg_{R}}{d\ln \mu}&=-\left(5+\frac{n_H}{2}\right)g_{R}^2\\
\left(4\pi \right)^2\frac{dg_{c}}{d\ln \mu}&=-5g_{c}^2\\
\end{split}
\end{equation}

\subsection{Yukawa couplings}

\begin{align}
\left(4\pi \right)^2\frac{dy_{\psi _{Q1}}}{d\ln \mu} &= y_{\psi _{Q1}}\left(-4g_L^2-4g_R^2-8g_c^2+2y_{\psi _{E1}}^2+6y_{\psi _{Q2}}^2\right)+2y_{\psi _{E1}}y_{\psi _{E2}}y_{\psi _{Q2}}+6y_{\psi _{Q1}}^3 \nonumber \\
\left(4\pi \right)^2\frac{dy_{\psi _{Q2}}}{d\ln \mu} &= y_{\psi _{Q2}}\left(-4g_L^2-4g_R^2-8g_c^2+2y_{\psi _{E2}}^2+6y_{\psi _{Q1}}^2\right)+2y_{\psi _{E1}}y_{\psi _{E2}}y_{\psi _{Q1}}+6y_{\psi _{Q2}}^3 \nonumber\\
\left(4\pi\right)^2\frac{d y_{\psi _{E1}}}{d\ln \mu} &= y_{\psi _{E1}}\left(-8g_L^2-8g_R^2+6y_{\psi _{E2}}^2+3y_{\psi _{Q1}}^2\right)+6y_{\psi _{E1}}^3+3y_{\psi _{E2}}y_{\psi _{Q1}}y_{\psi _{Q2}} \\
\left(4\pi\right)^2\frac{d y_{\psi _{E2}}}{d\ln \mu} &= y_{\psi _{E2}}\left(-8g_L^2-8g_R^2+6y_{\psi _{E1}}^2+3y_{\psi _{Q2}}^2\right)+6y_{\psi _{E2}}^3+3y_{\psi _{E1}}y_{\psi _{Q1}}y_{\psi _{Q2}} \nonumber
\end{align}

\subsection{Quartic couplings}

For single Higgs multiplet, $n_H=1$ and $y_{\psi _{E1}}:=y_{\psi _{E}}$, $y_{\psi _{Q1}}:=y_{\psi _{Q}}$, $y_{\psi _{E2}}=y_{\psi _{Q2}}=0$. The RGE for the quartic couplings are:
\begin{align}
(4\pi )^2\frac{d\lambda_a}{d\ln \mu} &= \lambda_a\left(-16g_L^2-16g_R^2+48\lambda_b+12y^2_{\psi_Q}+8y^2_{\psi_E}\right)\\
&+\frac{10}{3}g^2_L g^2_R+\frac{11g_L^4}{12}+\frac{11g^4_R}{12}+52\lambda_a^2+12\lambda_b^2-2y^4_{\psi_{E}}\\
(4\pi )^2\frac{d\lambda_b}{d\ln \mu} &= \lambda_b\left(-16g_L^2-16g_R^2+24\lambda_a+12y^2_{\psi_Q}+8y^2_{\psi_E}\right)\\
&-2g^2_L g^2_R+\frac{5g_L^4}{4}+\frac{5g^4_R}{4}+24\lambda_a^2-6y^4_{\psi_{Q}}-2y^4_{\psi_{E}}
\end{align}

For two Higgs multiplet, $n_H=2$. The RGE for the quartic couplings are:
\noindent
\begin{align}
(4\pi)^2\frac{d\lambda_{1a}}{d\ln \mu} &= \lambda_{1a}\left(48\lambda_{1b}-16g_L^2-16g_R^2+8y_{\psi_{E1}}^2+12y_{\psi_{Q1}}^2\right)+52\lambda_{1a}^2+12\lambda_{1b}^2\\  \nonumber
&+2\lambda_{a}\lambda_{b}+6\lambda_{a}\lambda_{c}+6\lambda_{a}\lambda_{d}+9\lambda_{a}^2+\lambda_{b}^2+2\lambda_{c}\lambda_{d}+4\lambda_{e}^2+4\lambda_{f}^2+\frac{10}{3}g_L^2 g_R^2\\ \nonumber
&+\frac{11}{12}g_L^4+\frac{11}{12}g_R^4-2y_{\psi_{E1}}^4\\ 
(4\pi)^2\frac{d\lambda_{1b}}{d\ln \mu} &=\lambda_{1b}\left(24\lambda_{1a}-16g_L^2-16g_R^2+8y_{\psi_{E1}}^2+12y_{\psi_{Q1}}^2\right)+24\lambda_{1b}^2+2\lambda_{b}\lambda_{c}\\\nonumber
&+2\lambda_{b}\lambda_{d}+3\lambda_{c}^2+3\lambda_{d}^2+8\lambda_{e}\lambda_{f}-2g_L^2 g_R^2+\frac{5}{4}g_L^4+\frac{5}{4}g_R^4-2y_{\psi_{E1}}^4-6y_{\psi_{Q1}}^4
\end{align}
\begin{align}
(4\pi)^2\frac{d\lambda_{2a}}{d\ln \mu} &=\lambda_{2a}\left(48\lambda_{2b}-16g_L^2-16g_R^2+8y_{\psi_{E2}}^2+12y_{\psi_{Q2}}^2\right)+52\lambda_{2a}^2+18\lambda_{2b}^2\\ \nonumber
&+2\lambda_{a}\lambda_{b}+6\lambda_{a}\lambda_{c}+6\lambda_{a}\lambda_{d}+9\lambda_{a}^2+\lambda_{b}^2+2\lambda_{c}\lambda_{d}+4\lambda_{e}^2+4\lambda_{f}^2\\ \nonumber
&+\frac{10}{3}g_L^2 g_R^2+\frac{11}{12}g_L^4+\frac{11}{12}g_R^4-2y_{\psi_{E2}}^4\\ 
(4\pi)^2\frac{d\lambda_{2b}}{d\ln \mu} &=\lambda_{2b}\left(24\lambda_{2a}-16g_L^2-16g_R^2+8y_{\psi_{E2}}^2+12y_{\psi_{Q2}}^2\right)+36\lambda_{2b}^2+2\lambda_{b}\lambda_{c}\\ \nonumber
&+2\lambda_{b}\lambda_{d}+3\lambda_{c}^2+3\lambda_{d}^2+8\lambda_{e}\lambda_{f}-2g_L^2 g_R^2+\frac{5}{4}g_L^4+\frac{5}{4}g_R^4-2y_{\psi_{E2}}^4-6y_{\psi_{Q2}}^4
\end{align}

For the scalar mixing couplings: 
\begin{align}
(4\pi)^2\frac{d\lambda_{a}}{d\ln \mu} &=\lambda_{a}\bigg(40\lambda_{1a}+24\lambda_{1b}+40\lambda_{2a}+24\lambda_{2b}-16g_L^2-16g_R^2+4y_{\psi_{E1}}^2+4y_{\psi_{E2}}^2\\\nonumber
&+6y_{\psi_{Q1}}^2+6y_{\psi_{Q2}}^2\bigg)+4\lambda_{1a}\lambda_{b}+12\lambda_{1a}\lambda_{c}+12\lambda_{1a}\lambda_{d}+4\lambda_{1b}\lambda_{c}+4\lambda_{1b}\lambda_{d}\\\nonumber
&+4\lambda_{2a}\lambda_{b}+12\lambda_{2a}\lambda_{c}+12\lambda_{2a}\lambda_{d}+4\lambda_{2b}\lambda_{c}+4\lambda_{2b}\lambda_{d}+2\lambda_{2b}^2+4\lambda_{a}^2+2\lambda_{b}^2\\\nonumber
&+2\lambda_{c}^2+2\lambda_{d}^2+8\lambda_{e}^2+\frac{2}{3}g_L^2 g_R^2+\frac{11}{6}g_L^4+\frac{11}{6}g_R^4-4y_{\psi_{E1}}^2y_{\psi_{E2}}^2\\
(4\pi)^2\frac{d\lambda_{b}}{d\ln \mu} &=2\lambda_{b}\bigg(2\lambda_{1a}+2\lambda_{2a}+9\lambda_{b}-8g_L^2-8g_R^2+2y_{\psi_{E1}}^2+2y_{\psi_{E2}}^2+3y_{\psi_{Q1}}^2\\ \nonumber
&+3y_{\psi_{Q2}}^2+4\lambda_{a}+6\lambda_{c}+6\lambda_{d}\bigg)+6\lambda_{2b}^2+8\lambda_{c}\lambda_{d}+96\lambda_{e}\lambda_{f}+88\lambda_{e}^2\\ \nonumber
&+16\lambda_{f}^2+ 6g_L^2 g_R^2-4y_{\psi_{E1}}^2y_{\psi_{E2}}^2\\
(4\pi)^2\frac{d\lambda_{c}}{d\ln \mu} &=\lambda_{c}\bigg(4\lambda_{1a}+12\lambda_{1b}+4\lambda_{2a}+12\lambda_{2b}+8\lambda_{a}-16g_L^2-16g_R^2+4y_{\psi_{E1}}^2\\\nonumber
&+4y_{\psi_{E2}}^2+6y_{\psi_{Q1}}^2+6y_{\psi_{Q2}}^2\bigg)+4\lambda_{1b}\lambda_{b}+4\lambda_{2b}\lambda_{b}+4\lambda_{b}\lambda_{d}+16\lambda_{e}\lambda_{f}\\\nonumber
&+6\lambda_{2b}^2+6\lambda_{c}^2+24\lambda_{f}^2-2g_L^2 g_R^2+\frac{5}{2}g_R^4-4y_{\psi_{E1}}^2y_{\psi_{E2}}^2-12y_{\psi_{Q1}}^2y_{\psi_{Q2}}^2\\
\end{align}
\begin{align}
(4\pi)^2\frac{d\lambda_{d}}{d\ln \mu} &=\lambda_{d}\bigg(4\lambda_{1a}+12\lambda_{1b}+4\lambda_{2a}+12\lambda_{2b}+8\lambda_{a}-16g_L^2-16g_R^2+4y_{\psi_{E1}}^2\\ \nonumber
&+4y_{\psi_{E2}}^2+6y_{\psi_{Q1}}^2+6y_{\psi_{Q2}}^2\bigg)+4\lambda_{1b}\lambda_{b}+4\lambda_{2b}\lambda_{b}+4\lambda_{b}\lambda_{c}+16\lambda_{e}\lambda_{f}\\ \nonumber
&+6\lambda_{2b}^2+6\lambda_{d}^2+24\lambda_{f}^2-2g_L^2 g_R^2+\frac{5}{2}g_L^4-4y_{\psi_{E1}}^2y_{\psi_{E2}}^2-12y_{\psi_{Q1}}^2y_{\psi_{Q2}}^2\\
(4\pi)^2\frac{d\lambda_{e}}{d\ln \mu} &=\lambda_{e}\bigg(4\lambda_{1a}+4\lambda_{2a}+8\lambda_{a}+40\lambda_{b}+12\lambda_{c}+12\lambda_{d}-16g_L^2-16g_R^2\\ \nonumber
&+4y_{\psi_{E1}}^2+4y_{\psi_{E2}}^2+6y_{\psi_{Q1}}^2+6y_{\psi_{Q2}}^2\bigg)+4\lambda_{1b}\lambda_{f}+4\lambda_{2b}\lambda_{f}+2\lambda_{2b}^2\\ \nonumber
&+24\lambda_{b}\lambda_{f}+4\lambda_{c}\lambda_{f}+4\lambda_{d}\lambda_{f}-2y_{\psi_{E1}}^2y_{\psi_{E2}}^2\\
(4\pi)^2\frac{d\lambda_{f}}{d\ln \mu} &=\lambda_{f}\bigg(4\lambda_{1a}+4\lambda_{2a}+8\lambda_{a}+12\lambda_{c}+12\lambda_{d}-16g_L^2-16g_R^2+4y_{\psi_{E1}}^2\\\nonumber
&+4y_{\psi_{E2}}^2+6y_{\psi_{Q1}}^2+6y_{\psi_{Q2}}^2\bigg)+4\lambda_{1b}\lambda_{e}+4\lambda_{2b}\lambda_{e}+6\lambda_{2b}^2+4\lambda_{c}\lambda_{e}+4\lambda_{d}\lambda_{e}\\\nonumber
&-2y_{\psi_{E1}}^2y_{\psi_{E2}}^2-6y_{\psi_{Q1}}^2y_{\psi_{Q2}}^2
\end{align}


\end{document}